%% file: TCOM19-Autoencoder.tex
\documentclass[journal]{IEEEtran}

\usepackage[nolist]{acronym} 
\usepackage{tikz}
\usepackage{pgfplots}
\usepgfplotslibrary{groupplots}
\usepackage{graphicx}
\usepackage{subcaption}
\pgfplotsset{compat=newest}
\usepackage{hyperref}
\usepackage{units}
\usepackage{amsmath, amsbsy, amssymb}
\usepackage{tikzscale}
\usepackage{color}
\usepackage{epigraph} 
\usepackage{circuitikz}
\usepackage{mathtools}
\usepackage[ruled,boxed,lined,linesnumbered]{algorithm2e}
\usepackage[group-separator={,}]{siunitx}
\usepackage{float}
\usepackage{pstricks}
\usepackage{dblfloatfix}    %

\input{corporateColours.tex}

\usetikzlibrary{shapes,arrows}
\usetikzlibrary{spy,backgrounds}
\tikzset{>=latex}

\usetikzlibrary{external}
\begin{document}

\title{Trainable Communication Systems:\\ Concepts and Prototype}

\author{Sebastian Cammerer, \textit{Student Member, IEEE}, Fay\c{c}al Ait Aoudia, \textit{Member, IEEE},\\ Sebastian D\"orner, \textit{Student Member, IEEE}, Maximilian Stark, \textit{Student Member, IEEE},\\ Jakob Hoydis, \textit{Senior Member, IEEE}, and Stephan ten Brink, \textit{Senior Member, IEEE}%
\\\vspace*{0.2cm}\textit{(Invited Paper)}

\thanks{S.~Cammerer, S.~D\"orner, and S.~ten~Brink are with the Institute of Telecommunications, University of  Stuttgart, Pfaffenwaldring 47, 70659 Stuttgart, Germany  (\{cammerer,doerner,tenbrink\}@inue.uni-stuttgart.de).
This work has been supported by DFG, Germany, under grant BR 3205/6-1.}%

\thanks{F.~Ait~Aoudia and J.~Hoydis are with Nokia Bell Labs, Route de Villejust, 91620 Nozay, France (\{faycal.ait\_aoudia,jakob.hoydis\}@nokia-bell-labs.com).}

\thanks{M.~Stark is with the Institute of Communications, Hamburg University of Technology, Hamburg, Germany (maximilian.stark@tuhh.de).}
}

\maketitle

\input{macros.tex}

\input{acronyms.tex}

\begin{abstract}
\input{tex/abstract}
\end{abstract}

\acresetall

\section{Introduction}

\input{tex/introduction}

\section{Bit-wise autoencoder} \label{sec:ae-nid}

\input{tex/ae_framework}

\section{Bit-wise iterative autoencoder} \label{sec:ae-id}

\input{tex/it_ae_framework}

\section{Code optimization} \label{sec:cd}

\input{tex/code_opt}

\section{Over-the-air experiments} \label{sec:ota}

\input{tex/ota.tex}

\section{Conclusion}\label{sec:conclu}

\input{tex/conclu.tex}

\section*{Acknowledgments}

\input{tex/acknowledgment}

\bibliographystyle{IEEEtran}
\bibliography{IEEEabrv,references}

\end{document}

%% file: corporateColours.tex
\definecolor{mittelblau}{RGB}{0, 126, 198}
\definecolor{violettblau}{cmyk}{0.9, 0.6, 0, 0}
\definecolor{rot}{RGB}{238, 28 35}
\definecolor{apfelgruen}{RGB}{140, 198, 62}
\definecolor{gelb}{RGB}{1, 221, 0}
\definecolor{orange}{RGB}{244, 111, 33}
\definecolor{pink}{RGB}{237, 0, 140}
\definecolor{lila}{RGB}{128, 10, 145}
\definecolor{hellgrau}{RGB}{224, 224, 224}
\definecolor{mittelgrau}{RGB}{128, 128, 128}
\definecolor{dunkelgrau}{RGB}{80,80,80}
\definecolor{anthrazit}{RGB}{19, 31, 31}

%% file: macros.tex
\renewcommand{\vec}[1]{\mathbf{#1}}
\newcommand{\vecs}[1]{\boldsymbol{#1}}

\newcommand{\av}{\vec{a}}
\newcommand{\bv}{\vec{b}}
\newcommand{\cv}{\vec{c}}
\newcommand{\dv}{\vec{d}}
\newcommand{\ev}{\vec{e}}
\newcommand{\fv}{\vec{f}}
\newcommand{\gv}{\vec{g}}
\newcommand{\hv}{\vec{h}}
\newcommand{\iv}{\vec{i}}
\newcommand{\jv}{\vec{j}}
\newcommand{\kv}{\vec{k}}
\newcommand{\lv}{\vec{l}}
\newcommand{\mv}{\vec{m}}
\newcommand{\nv}{\vec{n}}
\newcommand{\ov}{\vec{o}}
\newcommand{\pv}{\vec{p}}
\newcommand{\qv}{\vec{q}}
\newcommand{\rv}{\vec{r}}
\newcommand{\sv}{\vec{s}}
\newcommand{\tv}{\vec{t}}
\newcommand{\uv}{\vec{u}}
\newcommand{\vv}{\vec{v}}
\newcommand{\wv}{\vec{w}}
\newcommand{\xv}{\vec{x}}
\newcommand{\yv}{\vec{y}}
\newcommand{\zv}{\vec{z}}
\newcommand{\zerov}{\vec{0}}
\newcommand{\onev}{\vec{1}}
\newcommand{\alphav}{\vecs{\alpha}}
\newcommand{\betav}{\vecs{\beta}}
\newcommand{\gammav}{\vecs{\gamma}}
\newcommand{\lambdav}{\vecs{\lambda}}
\newcommand{\omegav}{\vecs{\omega}}
\newcommand{\sigmav}{\vecs{\sigma}}
\newcommand{\tauv}{\vecs{\tau}}
\newcommand{\thetav}{\vecs{\theta}}

\newcommand{\Am}{\vec{A}}
\newcommand{\Bm}{\vec{B}}
\newcommand{\Cm}{\vec{C}}
\newcommand{\Dm}{\vec{D}}
\newcommand{\Em}{\vec{E}}
\newcommand{\Fm}{\vec{F}}
\newcommand{\Gm}{\vec{G}}
\newcommand{\Hm}{\vec{H}}
\newcommand{\Id}{\vec{I}}
\newcommand{\Jm}{\vec{J}}
\newcommand{\Km}{\vec{K}}
\newcommand{\Lm}{\vec{L}}
\newcommand{\Mm}{\vec{M}}
\newcommand{\Nm}{\vec{N}}
\newcommand{\Om}{\vec{O}}
\newcommand{\Pm}{\vec{P}}
\newcommand{\Qm}{\vec{Q}}
\newcommand{\Rm}{\vec{R}}
\newcommand{\Sm}{\vec{S}}
\newcommand{\Tm}{\vec{T}}
\newcommand{\Um}{\vec{U}}
\newcommand{\Vm}{\vec{V}}
\newcommand{\Wm}{\vec{W}}
\newcommand{\Xm}{\vec{X}}
\newcommand{\Ym}{\vec{Y}}
\newcommand{\Zm}{\vec{Z}}
\newcommand{\Lambdam}{\vecs{\Lambda}}
\newcommand{\Pim}{\vecs{\Pi}}

\newcommand{\Bc}{{\cal B}}
\newcommand{\Cc}{{\cal C}}
\newcommand{\Dc}{{\cal D}}
\newcommand{\Ec}{{\cal E}}
\newcommand{\Fc}{{\cal F}}
\newcommand{\Gc}{{\cal G}}
\newcommand{\Hc}{{\cal H}}
\newcommand{\Ic}{{\cal I}}
\newcommand{\Jc}{{\cal J}}
\newcommand{\Kc}{{\cal K}}
\newcommand{\Lc}{{\cal L}}
\newcommand{\Mc}{{\cal M}}
\newcommand{\Nc}{{\cal N}}
\newcommand{\Oc}{{\cal O}}
\newcommand{\Pc}{{\cal P}}
\newcommand{\Qc}{{\cal Q}}
\newcommand{\Rc}{{\cal R}}
\newcommand{\Sc}{{\cal S}}
\newcommand{\Tc}{{\cal T}}
\newcommand{\Uc}{{\cal U}}
\newcommand{\Wc}{{\cal W}}
\newcommand{\Vc}{{\cal V}}
\newcommand{\Xc}{{\cal X}}
\newcommand{\Yc}{{\cal Y}}
\newcommand{\Zc}{{\cal Z}}

\newcommand{\CN}{\Cc\Nc}

\newcommand{\CC}{\mathbb{C}}
\newcommand{\MM}{\mathbb{M}}
\newcommand{\NN}{\mathbb{N}}
\newcommand{\RR}{\mathbb{R}}

\newcommand{\htp}{^{\mathsf{H}}}
\newcommand{\tp}{^{\mathsf{T}}}

\newcommand{\LB}{\left(}
\newcommand{\RB}{\right)}
\newcommand{\LP}{\left\{}
\newcommand{\RP}{\right\}}
\newcommand{\LSB}{\left[}
\newcommand{\RSB}{\right]}

\renewcommand{\ln}[1]{\mathop{\mathrm{ln}}\LB #1\RB}
\newcommand\norm[1]{\left\lVert#1\right\rVert}
\newcommand{\cs}[1]{\mathop{\mathrm{cs}}\LSB #1\RSB}

\newcommand{\EE}{{\mathbb{E}}}
\newcommand{\Expect}[2]{\EE_{#1}\LSB #2\RSB}

\newtheorem{definition}{Definition}[section]
\newtheorem{remark}{Remark}

%% file: acronyms.tex
\begin{acronym}
 \acro{ADC}{analog-to-digital converter}
 \acro{AGC}{automatic gain control}
 \acro{ASIC}{application-specific integrated circuit}
 \acro{AWGN}{additive white Gaussian noise}
 \acro{BER}{bit error rate}
 \acro{BICM}{bit interleaved coded modulation}
 \acro{BLER}{block error rate}
 \acro{SER}{symbol error rate}
 \acro{CFO}{carrier frequency offset}
 \acro{DL}{deep learning}
 \acro{DQPSK}{differential quadrature phase-shift keying}
 \acro{FPGA}{field programmable gate array}
 \acro{GNR}{GNU Radio}
 \acro{GPU}{graphic processing unit}
 \acro{ISI}{inter-symbol interference}
 \acro{LOS}{line-of-sight}
 \acro{MIMO}{multiple-input multiple-output}
 \acro{ML}{machine learning}
 \acro{MLP}{multilayer perceptron}
 \acro{MSE}{mean squared error}
 \acro{NN}{neural network}
 \acro{PLL}{phase-locked loop}
 \acro{ppm}{parts per million}
 \acro{PSK}{phase-shift keying}
 \acro{PFB}{polyphase filterbank}
 \acro{QAM}{quadrature amplitude modulation}
 \acro{ReLU}{rectified linear unit}
 \acro{RNN}{recurrent neural network}
 \acro{RRC}{root-raised cosine}
 \acro{RTN}{radio transformer network}
 \acro{SDR}{software-defined radio}
 \acro{SFO}{sampling frequency offset}
 \acro{SGD}{stochastic gradient descent}
 \acro{SNR}{signal-to-noise ratio}	
 \acro{TDL}{tapped delay line}
 \acro{LLR}{log-likelihood ratio}
 \acro{AE}{autoencoder}
 \acro{BMD}{bit-metric decoding}
 \acro{SMD}{symbol-metric decoding}
 \acro{MI}{mutual information}
 \acro{IDD}{iterative demapping and decoding}
 \acro{ECC}{error correction code}
 \acro{FEC}{forward error correction}
 \acro{LDPC}{low-density parity-check}
 \acro{BP}{belief propagation}
 \acro{TX}{transmitter}
 \acro{RX}{receiver}
 \acro{EXIT}{extrinsic information transfer}
 \acro{CNN}{convolutional neural network}
 \acro{BICM-ID}{bit interleaved coded modulation with iterative demapping}
 \acro{VN}{variable node}
 \acro{VND}{variable node decoder}
 \acro{CN}{check node}
 \acro{CND}{check node decoder}
 \acro{DNN}{dense neural network}
 \acro{OTA}{over-the-air}
 \acro{MAP}{maximum a posteriori}
 \acro{EVM}{error vector magnitude}
 \acro{CE}{cross entropy}
 \acro{KL}{Kullback-Leibler}
 \acro{APSK}{amplitude and phase-shift keying}
 \acro{BMI}{bit-wise mutual information}
 \acro{SMI}{symbolwise mutual information}
 \acro{wrt}[w.r.t.\@]{with respect to} 
 \acro{i.i.d.}{independent and identically distributed}
 \acro{CP}{cyclic prefix}
 \acro{MMSE}{minimum mean square error}
 \acro{USRP}{Universal Software Radio Peripheral}
 \acro{OFDM}{Orthogonal Frequency-Division Multiplexing}
 \acro{GPU}{graphics processing unit}
 \acro{PEG}{progressive edge-growth}
 \acro{MCS}{modulation and coding scheme}
 \acro{NUC}{non-uniform constellation}
\end{acronym}

%% file: tex/abstract.tex
We consider a trainable point-to-point communication system, where both transmitter and receiver are implemented as  \acp{NN}, and demonstrate that training on the \ac{BMI} allows seamless integration with practical \ac{BMD} receivers, as well as joint optimization of constellation shaping and labeling.
Moreover, we present a fully differentiable neural \ac{IDD} structure which achieves significant gains on \ac{AWGN} channels using a standard 802.11n \ac{LDPC} code.
The strength of this approach is that it can be applied to arbitrary channels without any modifications.
Going one step further, we show that careful code design can lead to further performance improvements.
Lastly, we show the viability of the proposed system through implementation on \acp{SDR} and training of the end-to-end system on the actual wireless channel.
Experimental results reveal that the proposed method enables significant gains compared to conventional techniques.

\begin{IEEEkeywords}
Autoencoder, end-to-end learning, iterative demapping and decoding, code design, geometric shaping, software-defined radio
\end{IEEEkeywords}

%% file: tex/introduction.tex
End-to-end learning of communication systems consists of implementing the transmitter, channel, and receiver as a single \ac{NN}, referred to as an autoencoder, and training it to reproduce its input at its output~\cite{8054694}.
This approach enables joint optimization of the transmitter and receiver for a specific channel model without extensive mathematical analysis.
Autoencoder-based communication systems have first been proposed in the context of wireless communications~\cite{8054694}, and have subsequently been extended towards other settings, such as optical fiber~\cite{Karanov:18}, optical wireless~\cite{8819929}, and molecular communications~\cite{8715741}.
Most of these approaches are optimized on the symbol-wise categorical \ac{CE}, which is equivalent to maximizing the mutual information between the channel input and output~\cite{GC2019}.
However, practical systems usually rely on \ac{BICM}~\cite{669123} because of its reasonable complexity.
With \ac{BICM}, \ac{BMD}~\cite{7307154} is used at the receiver, in which the points forming the channel input constellation are labeled by bit vectors, and bit levels are treated independently.
Therefore, the points that form a constellation learned by an autoencoder optimized on the symbol-wise categorical \ac{CE} must be labeled by bit vectors to be used in a practical \ac{BMD}-based system.
This is typically a delicate task, even for low modulation orders.
Moreover, the mutual information between channel input and output is known to \emph{not} be a rate achievable by \ac{BMD}~\cite{7307154}.
This suggests that this metric is not appropriate for trainable communication systems leveraging \ac{BMD}.

\begin{figure*}[t]
 	\centering
  	\begin{subfigure}[b]{0.48\linewidth}
 		\centering
 		\tikzsetnextfilename{symbol_ae_setup}
		\resizebox{1.0\linewidth}{!}{\includegraphics{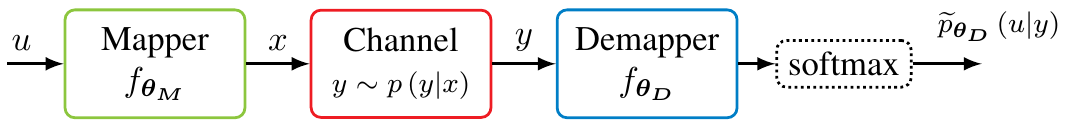}}
		\caption{Symbol-wise autoencoder\label{fig:symbolwise-ae}}
	\end{subfigure} \quad
 	\begin{subfigure}[b]{0.48\linewidth}
 		\centering
 		\tikzsetnextfilename{bit_ae_setup}
		\resizebox{1.0\linewidth}{!}{\includegraphics{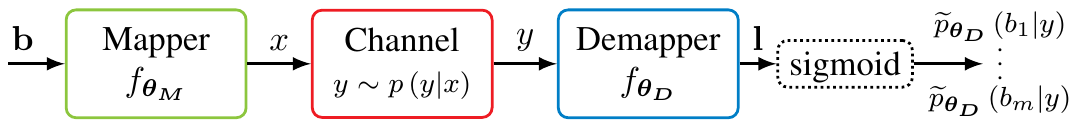}}
		\caption{Bit-wise autoencoder \label{fig:bitwise-ae}}
	\end{subfigure}
	\caption{Symbol-wise and bit-wise autoencoder architectures}
	\vspace*{-0.3cm}
\end{figure*}

In this work, training on the \ac{BMI} is considered instead.
The \ac{BMI} was proven to be an achievable rate for \ac{BMD}~\cite{bocherer14}, making it a suitable metric for the optimization of communication systems based on \ac{BMD}.
Moreover, training on the \ac{BMI} enables joint optimization of constellation shaping and labeling on the transmitter side.
On the receiver side, the learned demapper outputs \acp{LLR} and can therefore be smoothly interfaced with a channel decoder, e.g., based on \ac{BP}. Since the learned constellations benefit from \ac{IDD}~\cite{775793}, we present a differentiable neural \ac{IDD} architecture which unlocks significant gains on the \ac{AWGN} channel, considering a standard 802.11n \ac{LDPC} code.
We then show how further performance improvements can be achieved by designing an \ac{LDPC} code ``on top'' of the trained communication system.
We follow the basic idea in \cite{alberge2018deep} which describes, to the best of our knowledge, the first autoencoder with a training procedure that directly optimizes the bit-metrics.
A similar \emph{neural receiver structure} has been proposed in \cite{he2019robust} for visible light communications and, recently, in \cite{2019arXiv191110131K} for the optical fiber. While the authors of \cite{he2019robust} train the system for a symbol-wise output (and calculate the \acp{LLR} \emph{manually}), the authors of \cite{2019arXiv191110131K} use a bit-wise \emph{neural equalizer} and show performance gains over the optical fiber by a tailored \ac{LDPC} code. However, although both schemes consist of a trainable \ac{IDD} receiver, they both do not train the system end-to-end, i.e., no training at transmitter is involved.

Significant gains by using \acp{NUC} for the \ac{BICM} scenario have been shown in \cite{loghin2016non} based on \emph{conventional} optimization techniques. In \cite{alberge2018deep}, a \ac{BICM} autoencoder-based system for the \ac{AWGN} channel with additive radar interference is presented. It was also observed that the concatenation with an outer channel code simplifies if \ac{BMD}-based autoencoders are used and further improves the overall performance. %
In this work, we aim to put the bit-metric optimized autoencoder on a theoretical solid ground and introduce the bit-wise autoencoder in the \ac{BICM} scenario. %
Although all simulations in this paper were carried out considering an \ac{AWGN} channel, the proposed approach can be applied without modifications to any channel model.
Thus, we demonstrate the universality of the proposed method through its implementation on \acp{SDR}, and by training on an actual wireless channel, using the  algorithm proposed in~\cite{8792076} which enables training of the end-to-end system even if the channel gradient is not accessible or not defined.
The observed gains confirm the benefits of trainable communication systems in practice. Our learning-based optimization of the full physical layer could be completely automated and might hence be one of the key ingredients of next generation communication systems.

The rest of this paper is organized as follows.
Section~\ref{sec:ae-nid} motivates training on \ac{BMI} from an information-theoretical viewpoint.
The neural \ac{IDD} structure is introduced in Section~\ref{sec:ae-id}.
In Section~\ref{sec:cd}, code design is performed to achieve further gains, and Section~\ref{sec:ota} presents the results of over-the-air experiments.
Finally, Section~\ref{sec:conclu} concludes the paper.

\paragraph*{Notations}
Random variables are denoted by capital italic font, e.g., $X ,Y$, with realizations $x, y$, respectively.
$I(X;Y)$, $p(y|x)$ and $p(x,y)$ represent the mutual information, conditional probability, and joint probability distribution of the two random variables $X$ and $Y$. 
Multivariate random variables are represented with capital bold font, e.g., $\mathbf{Y} = [Y_0,Y_1]^T$.
Vectors are represented using a lower case bold font, e.g., $\mathbf{y}$.
The cross entropy for two discrete distributions $p(x)$ and $q(x)$ is $H(p(x), q(x)) \coloneqq -\sum_x p(x) \log{q(x)}$.

%% file: tex/ae_framework.tex
\begin{figure*}
 	\centering
 	\begin{subfigure}[b]{0.56\linewidth}
 		\centering
 		\tikzsetnextfilename{ae_setup}
		\resizebox{1.0\linewidth}{!}{\includegraphics{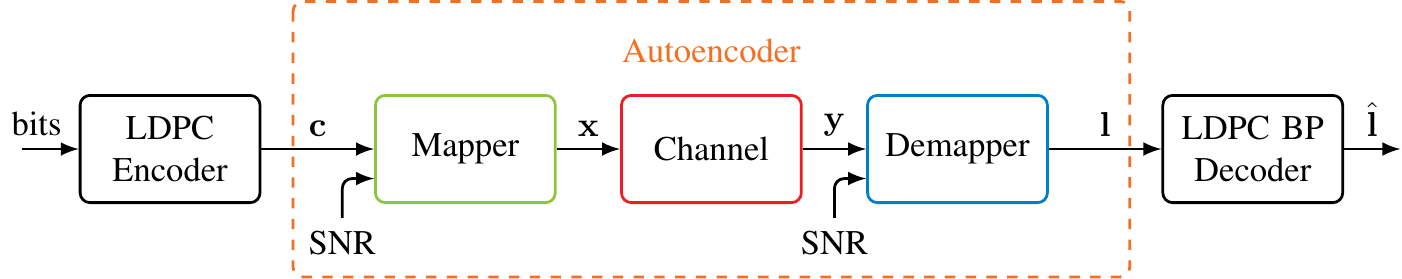}}
		\caption{Setup overview \label{fig:setup-nid}}
	\end{subfigure} \quad
 	\begin{subfigure}[b]{0.23\linewidth}
 		\centering
 		\tikzsetnextfilename{mapper_structure}
		\resizebox{1.0\linewidth}{!}{\includegraphics{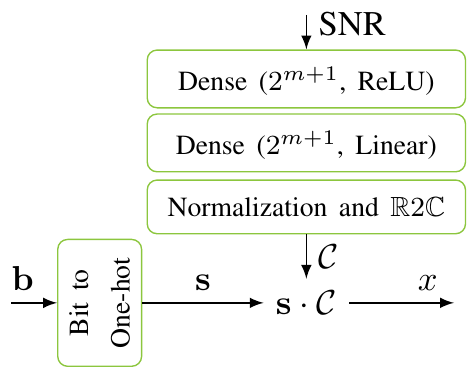}}
		\caption{Mapper\label{fig:mod-archi}}
	\end{subfigure} \quad
 	\begin{subfigure}[b]{0.15\linewidth}
 		\centering
 		\tikzsetnextfilename{demapper_structure}
		\resizebox{1.0\linewidth}{!}{\includegraphics{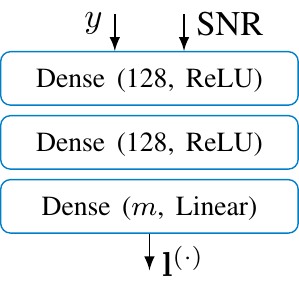}}
		\caption{Demapper\label{fig:demod-archi}}
	\end{subfigure}
	\caption{Bit-wise autoencoder setup with channel coding, where $\cv = \LSB \bv^{(1)\mathsf{T}},\dots,\bv^{(s)\mathsf{T}} \RSB\tp$ and $\lv = \LSB \lv^{(1)\mathsf{T}},\dots,\lv^{(s)\mathsf{T}} \RSB \tp$. For simplicity, the interleaver/deinterleaver are considered part of the \ac{LDPC} graph, and are therefore not shown.}
\end{figure*}

Although it appears as a trivial modification at first glance, the transition from symbol-wise to bit-wise autoencoder-based communication systems turns out to require a carefully adjusted framework.
The main difference in the objective is to minimize the \ac{BER} instead of the \ac{SER}.
This immediately leads to the question of how to find the optimal labeling scheme for the learned constellations, which is not part of previous autoencoder implementations.

Fig.~\ref{fig:symbolwise-ae} shows the architecture of such a typical autoencoder-based end-to-end communication system optimized on the symbol-wise categorical \ac{CE}.
For clarity, complex-unidimensional constellations are considered, as generalization to multidimensional constellations is trivial. 
The number of bits per channel use is denoted by $m$.
The mapper takes as input a message $u \in \{0,\dots,2^m-1\}$, and maps it to a complex baseband symbol $x$.
Therefore, it implements the mapping $f_{\boldsymbol{\theta_M}} : \{0,\dots,2^m-1\} \mapsto \CC$, with trainable parameters $\boldsymbol{\theta_M}$.
The complex baseband symbol is sent over a channel $y \sim p(y|x)$, with $y \in \CC$.
The received sample is fed to the demapper, which implements the mapping $g_{\boldsymbol{\theta_D}} : \CC \mapsto \RR^{2^m}$ with trainable parameters $\boldsymbol{\theta_D}$.
The $2^m$ outputs of the demapper are called \emph{logits}.
Probabilities over the $2^m$ messages, denoted by $\widetilde{p}_{\boldsymbol{\theta_D}}(u|y)$, are obtained by applying the softmax function to the logits. 
For a general introduction of the concept of end-to-end learning we refer the interested reader to \cite{8054694}.

The constellation obtained by training on the symbol-wise \ac{CE} needs to be labeled to be used in a conventional \ac{BICM} system. However, since the learned constellation points do not usually form a grid (see Fig.~\ref{fig:const-nid-16}), as in a conventional QAM, finding the optimal labeling is a combinatorial problem with $2^m!$ possibilities (neglecting symmetries). Therefore, we can only rely on sub-optimal heuristics to label the constellation points \emph{after} the training process, as, e.g., done in \cite{Karanov:18}.

\subsection{Information theory perspective}

An autoencoder-based communication system is typically trained by minimizing the categorical \ac{CE} between the true posterior distribution $p_{\boldsymbol{\theta_M}}(u|y)$\footnote{The posterior depends on the trainable parameters $\boldsymbol{\theta_M}$ of the mapper.} and the one learned by the receiver $\widetilde{p}_{\boldsymbol{\theta_D}}(u|y)$, averaged over all the possible channel outputs $y$, i.e.,
\begin{multline}
	\label{eq:cce}
	\Expect{y}{H \LB p_{\boldsymbol{\theta_M}}(u|y),\widetilde{p}_{\boldsymbol{\theta_D}}(u|y) \RB}\\
		= -\Expect{y}{\sum_{u=0}^{2^m-1} p_{\boldsymbol{\theta_M}}(u|y) \log{\widetilde{p}_{\boldsymbol{\theta_D}}(u|y)}}.
\end{multline}
Assuming that the \ac{NN} implementing the demapper has a sufficiently high capacity (i.e., is large enough), this is equivalent to maximizing the mutual information between the channel input $X$ and output $Y$~\cite{GC2019}.
It is known~\cite{7307154} that achieving the maximum $I(X;Y)$ requires \emph{multistage decoding} at the receiver, together with \emph{multilevel coding} at the transmitter, i.e., the use of an individual binary code on each bit level.
As a consequence, most practical systems rely on the simpler \ac{BICM} encoder~\cite{669123}, with which all bit-levels are encoded by a single binary code.
At the receiver, \ac{BMD} is used, which treats the bit levels independently.
For practical use of a constellation learned by an autoencoder trained on (\ref{eq:cce}) with \ac{BMD}, the constellation needs to be labeled, i.e., each constellation point needs to be mapped to a unique bit vector.
This task is non-trivial, and becomes quickly intractable even for small modulation orders.
Moreover, $I(X;Y)$ is not an achievable rate by \ac{BMD}~\cite{7307154}.
Therefore, even if an optimal labeling is assumed (e.g., found by exhaustive search), a constellation learned by minimization of (\ref{eq:cce}) is not necessarily optimal for practical \ac{BMD} receivers.

An achievable rate by \ac{BMD} is the \ac{BMI}~\cite{bocherer14}
\begin{equation}
	R \coloneqq H(\Bm) - \sum_{j=1}^m H(B_j|Y) \leq I(X;Y)
\end{equation}
where $\Bm$ denotes the binary random variable associated with the input bit vector $\bv$ of length $m$.
This result motivates the bit-wise autoencoder shown in Fig.~\ref{fig:bitwise-ae}.
In the bit-wise autoencoder, the mapper takes as input a bit vector of length $m$, and the demapper outputs $m$ logits (one per bit) which form a vector denoted by $\lv \in \RR^m$.
Probabilities over the $m$ bits, denoted by $\widetilde{p}_{\boldsymbol{\theta_D}}(b_j|y),~j=1,\dots,m$, are obtained by element-wisely applying the sigmoid function
$$\sigma(x) = \frac{1}{1+e^{-x}}$$
 to the corresponding logits.
One can see that the logits correspond to \acp{LLR}\footnote{Note that, we define the logits such that $\widetilde{p}_{\boldsymbol{\theta_D}}(b_j = 0|y) = \sigma\left(l_j\right)$ instead of the more common definition $\widetilde{p}_{\boldsymbol{\theta_D}}(b_j = 1|y) = \sigma\left(l_j\right)$. However, this only impacts the sign but allows to use logits and \acp{LLR} equally throughout this paper}:
\begin{align}
	\forall j \in \{1,\dots,m\},~\text{LLR}(j) &\coloneqq \log{\frac{1-\widetilde{p}_{\boldsymbol{\theta_D}}(b_j = 1|y)}{\widetilde{p}_{\boldsymbol{\theta_D}}(b_j = 1|y)}}\\
				  & = l_j
\end{align}
where
\begin{equation}
\widetilde{p}_{\boldsymbol{\theta_D}}(b_j = 0|y) = \sigma\left(l_j\right).
\end{equation}

Optimization of the bit-wise autoencoder is done by minimizing the total binary \ac{CE}
\begin{align}
	\Lc(\boldsymbol{\theta_M}, \boldsymbol{\theta_D}) &\coloneqq \sum_{j=1}^m \Expect{y}{H \LB p_{\boldsymbol{\theta_M}}(b_j|y),\widetilde{p}_{\boldsymbol{\theta_D}}(b_j|y) \RB } \label{eq:loss-nid}\\
		&= \sum_{j=1}^m \Expect{y,b_j}{-\log{\widetilde{p}_{\boldsymbol{\theta_D}}(b_j|y)}}
\end{align}
which is closely related to the \ac{BMI}
\begin{multline}
	\label{eq:bce}
	\Lc(\boldsymbol{\theta_M}, \boldsymbol{\theta_D}) = H(\Bm) - R\\
		+ \sum_{j=1}^m \Expect{y}{\text{D}_{\text{KL}} \LB p_{\boldsymbol{\theta_M}}(b_j|y)||\widetilde{p}_{\boldsymbol{\thetav_D}}(b_j|y)} \RB
\end{multline}
where $\text{D}_{\text{KL}}$ is the \ac{KL} divergence, and $p_{\boldsymbol{\theta_M}}(b_j|y),~j=1,\dots,m$, are the true posterior distributions. 
Interestingly, according to \cite{bocherer14}, \eqref{eq:loss-nid} itself is an achievable rate for an imperfect receiver. Rewriting \eqref{eq:loss-nid} as \eqref{eq:bce} allows to connect the actual \ac{BMI} to the achievable rate for imperfect receivers. In turn, it becomes clear how trainable communication systems are able to easily outperform conventional systems with imperfect receivers.
In more detail, the first term on the right-hand side of~(\ref{eq:bce}) is the entropy of the bit vector generated by the source, which is typically constant and equals the amount of bits $m$ mapped to one symbol (assuming uniformly distributed bits).
The second term is the \ac{BMI}, and the third term is the sum of the \ac{KL} divergences between the true posterior distributions $p_{\boldsymbol{\theta_M}}(b_j|y)$ and the ones learned by the demapper $\widetilde{p}_{\boldsymbol{\thetav_D}}(b_j|y)$, and accounts for the sub-optimality of the receiver.
Training of the end-to-end system by minimizing $\Lc$ therefore corresponds to maximizing $R$, which is suited to \ac{BMD}, while minimizing the \ac{KL} divergence between the optimal demapper and the one learned at the receiver.
Moreover, because the mapper assigns each bit vector to a constellation point, joint geometric shaping and bit labeling is performed when minimizing $\Lc$.
The \ac{NN} implementing the demapper should approximate the posterior distributions $p_{\boldsymbol{\theta_M}}(b_j|y)$ of a constellation maximizing the \ac{BMI} with high precision.
This avoids learning a constellation where the posterior distributions are well-approximated, but the \ac{BMI} is not maximized. This can be ensured by choosing the \ac{NN} implementing the demapper large enough, so that it is (in principle) capable of approximating a wide range of posterior distributions. Unfortunately, even with a very large \ac{NN}, there is no guarantee that the learning procedure converges to such a solution for every channel.

\subsection{Simulation setup} \label{sec:nid-setup}

We will now evaluate the autoencoder on an \ac{AWGN} channel with two-sided noise-power spectral-density $N_0$, i.e.,
\begin{equation}
	p(y|x) = \frac{1}{\pi N_0}\exp{ \LB -\frac{|y-x|^2}{N_0} \RB}.
\end{equation}
Fig.~\ref{fig:setup-nid} shows the end-to-end system.
A bitstream is fed to an \ac{LDPC} encoder which generates codewords $\cv$ of length $n$.
The number of bits per channel use is $m$ and it is assumed that $n$ is a multiple of $m$.
Each codeword is broken apart into $s = \frac{n}{m}$ bit vectors $\bv$ of length $m$, i.e., $\cv = \LSB \bv^{(1)\mathsf{T}},\dots,\bv^{(s)\mathsf{T}} \RSB \tp$.
Each bit vector $\bv^{(i)}, i=1,\dots,s$, is mapped into a complex baseband symbol $x_i \in \CC$, and is sent over the channel.
On the receiver side, the demapper processes each received sample $y_i \in \CC$ and generates \acp{LLR} $\lv^{(i)} \in \RR^m$.
Finally, the \acp{LLR} of the entire codeword $\lv = \LSB \lv^{(1)\mathsf{T}},\dots,\lv^{(s)\mathsf{T}} \RSB \tp$ are fed into a \ac{BP} decoder.
For simplicity, the interleaver/deinterleaver are considered part of the \ac{LDPC} graph, and are therefore not shown.

The architectures of the \acp{NN} that implement the mapper and the demapper are shown in Fig.~\ref{fig:mod-archi} and Fig.~\ref{fig:demod-archi}, respectively.
The mapper includes an \ac{NN} that generates a continuum of constellations $\Cc \in \CC^{2^m}$ that are determined by the \ac{SNR}.
Note that $\Cc$ determines the whole constellation set, i.e., contains $2^m$ complex-valued symbols and the one-hot vector $\sv$ selects the corresponding symbol.
The \ac{NN} included in the mapper is made of two dense layers of $2^{m+1}$ units each, one with \ac{ReLU} activations and the other with linear activations.
The $2^{m+1}$ outputs of the second dense layer correspond to the real and imaginary parts of the $2^m$ constellation points.
The last layer is a normalization layer, which ensures that $\Expect{x}{|x|^2} = 1$.
A bit vector $\bv$ is mapped to one of the constellation points.
To that aim, it is first converted to its one-hot representation denoted by $\sv$, i.e., the vector of size $2^m$ with all elements set to zero, except the one whose index has $\bv$ as binary representation set to one.
The vector $\bv$ is then mapped to a constellation point $x$ by taking the product of $\sv$ and $\Cc$.
Note that this is only the training procedure and the later implementation of the mapper can be further optimized in terms of implementation complexity.
The demapper is made of three dense layers, two with \ac{ReLU} activations and 128 units each, and the last one with linear activations and $m$ units.
Moreover, the demapper takes as input the \ac{SNR} (in dB).
This was motivated by the observation that the posterior distribution depends on the \ac{SNR}.

We want to emphasize that the constellation set in our setup is \ac{SNR} dependent, i.e., mapper and demapper both need to know the \ac{SNR}. However, this is just a generalization of the case when re-training per \ac{SNR} is performed as for trainable systems the \ac{SNR} is always implicitly part of the training data. In case that feedback of the \ac{SNR} is not possible, the same setup can be trained with fixed \ac{SNR} input at the price of a slightly reduced \ac{BER} performance. Recall that in \emph{classical} communication systems (e.g., in the 5G standard), we typically define the \ac{MCS} and, thus, the constellations (and accordingly the LDPC code) can be also adapted.

An 802.11n~\cite{5307322} irregular \ac{LDPC} code was considered, with rate $r = \frac{1}{2}$ and codewords of length $n = 1296\:$bit.
The \ac{SNR} is defined as
\begin{equation}
	\text{SNR} = \frac{1}{r m N_0}
\end{equation}
which corresponds to the energy per information-bit per noise-power spectral-density ratio since the normalization ensures $\Expect{x}{|x|^2} = 1$.
Training was done with the Adam~\cite{kingma2014} optimization algorithm and we have initialized the weights of the mapper using the \emph{Glorot initializer} \cite{glorot2010understanding}.
We have summarized the training parameters in Table~\ref{tab:training_param}.

The loss function $\Lc$ was estimated by
\begin{multline}
	\Lc(\boldsymbol{\theta_M}, \boldsymbol{\theta_D}) \approx -\frac{1}{|\Bc|}\sum_{\bv\in\Bc} \sum_{j=1}^m \Bigg( b_j \log{\LB \widetilde{p}_{\boldsymbol{\thetav_D}} \LB b_j|y \RB\RB}\\
		+ \LB 1 - b_j \RB \log{\LB 1 - \widetilde{p}_{\boldsymbol{\thetav_D}} \LB b_j|y \RB \RB} \Bigg)
\end{multline}
where $\Bc$ is a finite set of samples of bit vectors.
The \emph{batch-size} $|\Bc|$, i.e., the number of samples used to estimate $\Lc$, was set to 500, whereas the learning rate was progressively decreased from $10^{-3}$ to $10^{-5}$.
At training, the \ac{SNR} was randomly and uniformly selected for each example from a $4\:$dB range centered on the waterfall region of the code.
Note that, at training, the channel encoder and decoder are not needed, as the loss function~(\ref{eq:loss-nid}) is based on the output of the demapper.
At evaluation, the number of iterations performed by the \ac{BP} decoder was set to 40.

\begin{table}[]
\centering
\caption{Training Parameters}
\begin{tabular}{l l}
Parameter & Value\\
\hline
Batch-size & 500 \\
Learning rate & decreasing from $10^{-3}$ to  $10^{-5}$ \\
Training SNR & 4 dB around waterfall region of LDPC code\\
Optimizer & Adam \cite{kingma2014} \\
Initializer & Glorot \cite{glorot2010understanding}
\end{tabular}
\label{tab:training_param}
\end{table}

\subsection{Results}

\input{figs/ber_nid.tex}

\begin{figure*}
 	\centering
  	\begin{subfigure}[b]{\linewidth}
		\centering
		\tikzsetnextfilename{bit_ae_constellations}
		\includegraphics{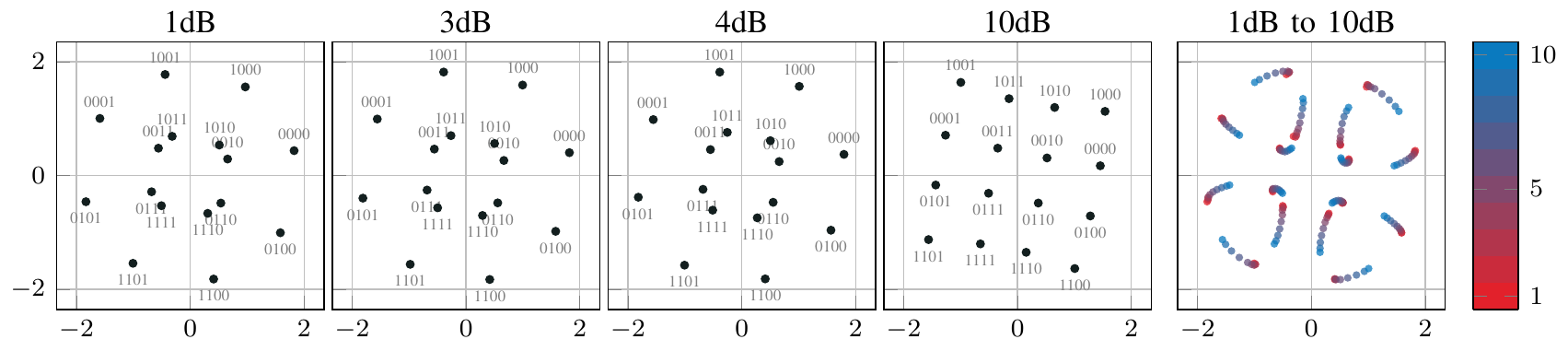}
		\caption{Bit-wise \label{fig:const-nid-bw}}
	\end{subfigure} \\ %
 	\begin{subfigure}[b]{\linewidth}
 		\centering
 		\tikzsetnextfilename{symbol_ae_constellations}
		\includegraphics{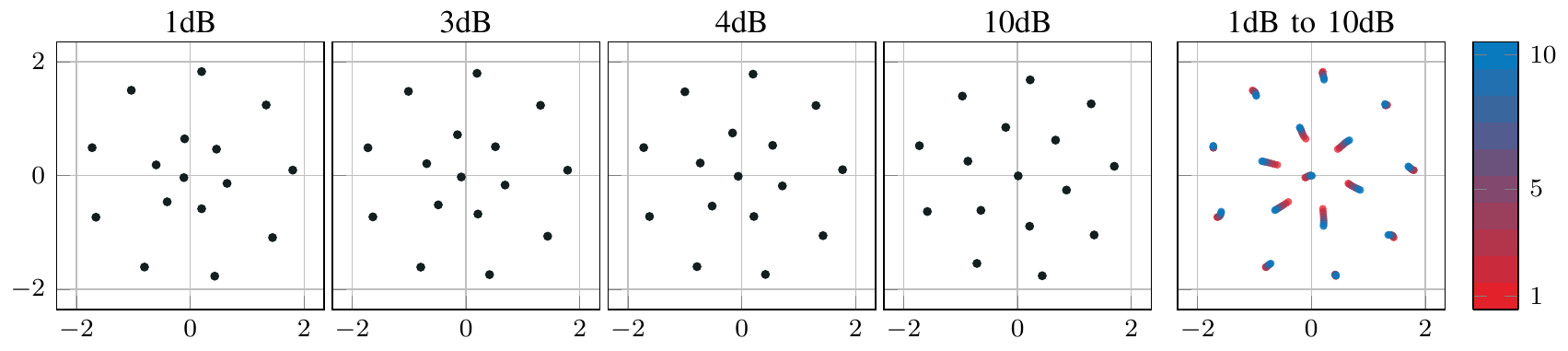}
		\caption{Symbol-wise \label{fig:const-nid-sw}}
	\end{subfigure}
	\caption{Constellations learned for $m=4$ bit at varying SNR. The learned labeling of the bit-wise autoencoder is shown in gray.\label{fig:const-nid-16}} %
\end{figure*}

\begin{figure}
  	\centering
  	\tikzsetnextfilename{symbol_ae_ber_nid}
	\includegraphics{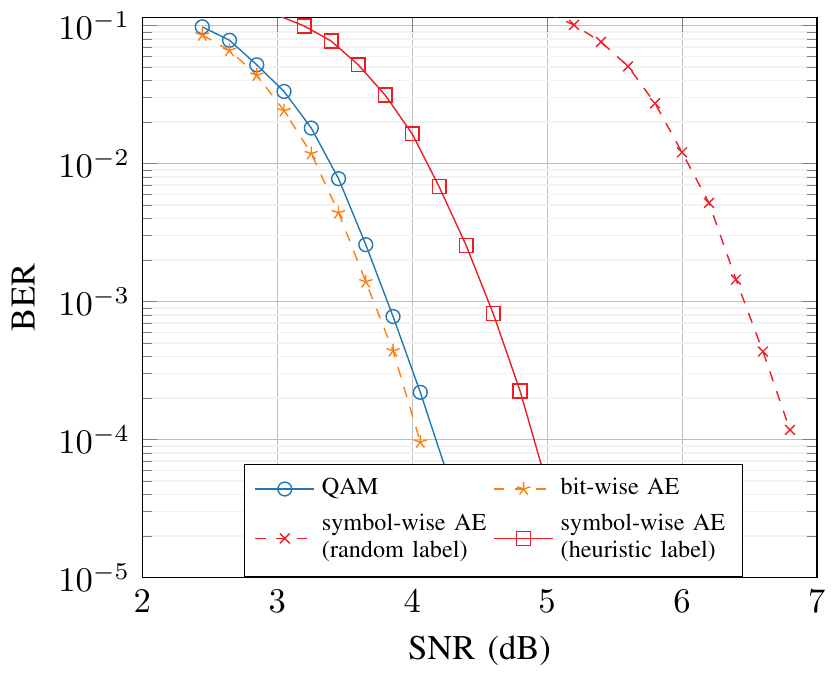}
	\caption{Achieved BER with $m=4$ bit for baseline \ac{QAM}, bit-wise autoencoder and symbol-wise autoencoder (with random and heuristic labeling). \label{fig:ber-symbolae}}
\end{figure}

The (coded) \ac{BER} achieved by the bit-wise autoencoder was compared to that of \ac{PSK} for $m=3$ and to \ac{QAM} for $m=4$,~$6$, and~$8$.
For the considered baselines, maximum-likelihood demapping was used.
Fig.~\ref{fig:ber-nid} shows the \ac{BER} achieved by the compared approaches with the setup presented in Section~\ref{sec:nid-setup}.
It can be seen that the schemes learned by the bit-wise autoencoder outperform the baselines.
Moreover, the gains achieved by the learning schemes increase with the modulation order, reaching $0.8\:$dB for $m=8$, compared to $0.3\:$dB for $m=3$.

To get insight into the learned constellation geometries, Fig.~\ref{fig:const-nid-16} shows the constellations learned for $m=4$, when leveraging the bit-wise and symbol-wise autoencoder, respectively.
It can be seen that training on the \ac{BMI} (Fig.~\ref{fig:const-nid-bw}) leads to a constellation which differs significantly from the one obtained from training on the symbol-wise \ac{CE} (Fig.~\ref{fig:const-nid-sw}). Recall that the use of the constellation learned with the symbol-wise autoencoder (Fig.~\ref{fig:const-nid-sw}) in a \ac{BMD}-based system requires an additional heuristic labeling step, which is typically not trivial.

Besides the labeling, the optimal position of the constellation points can differ under different metrics. An intuitive example is depicted in Fig.~\ref{fig:const-nid-16} where, for the low \ac{SNR} region, the bit-wise optimized autoencoder (Fig.\ref{fig:const-nid-bw}) \emph{clusters} constellation points into groups that only differ in one bit position. This effectively weakens the reliability of this bit position while improving the other positions and, thereby, optimizes the overall achievable information rate. However, in the corresponding symbol-metric a confusion is unavoidable within the clustered symbols and, therefore, a degraded symbol-wise performance can be observed. When looking at the corresponding symbol-metric optimized constellation in Fig.\ref{fig:const-nid-sw}, we see that such a clustering does not occur for the exact same channel parameters after training. While a symbol-wise trained constellation improves the \ac{SER}, it degrades the achievable \ac{BER}. %
Fig.~\ref{fig:ber-symbolae} shows the \ac{BER} performance of a symbol-wise trained autoencoder system with randomly chosen labels and heuristic labels compared with a \ac{QAM} baseline with Gray labeling and a bit-wise optimized autoencoder. We want to emphasize that our heuristic labeling is not necessarily optimal, but underlines the difficulties in finding such a labeling. As it can be seen, the \ac{BER} performance of the symbol-wise optimized autoencoder with heuristic labels is $\approx 0.7$dB worse than the conventional \ac{QAM} baseline and even $\approx 0.8$dB worse than the bit-wise optimized autoencoder. %
These results suggest that one should train on the \ac{BMI} when using \ac{BMD}.
Moreover, in the case of the bit-wise autoencoder, the learned labeling is also shown in Fig.~\ref{fig:const-nid-bw}.
One can see that a form of Gray labeling was learned, where two points close to each other are assigned labels that differ in only one bit.

%% file: figs/ber_nid.tex
\newlength{\rx}
\newlength{\ry}

\begin{figure}
		\tikzsetnextfilename{ber-nid}
\includegraphics{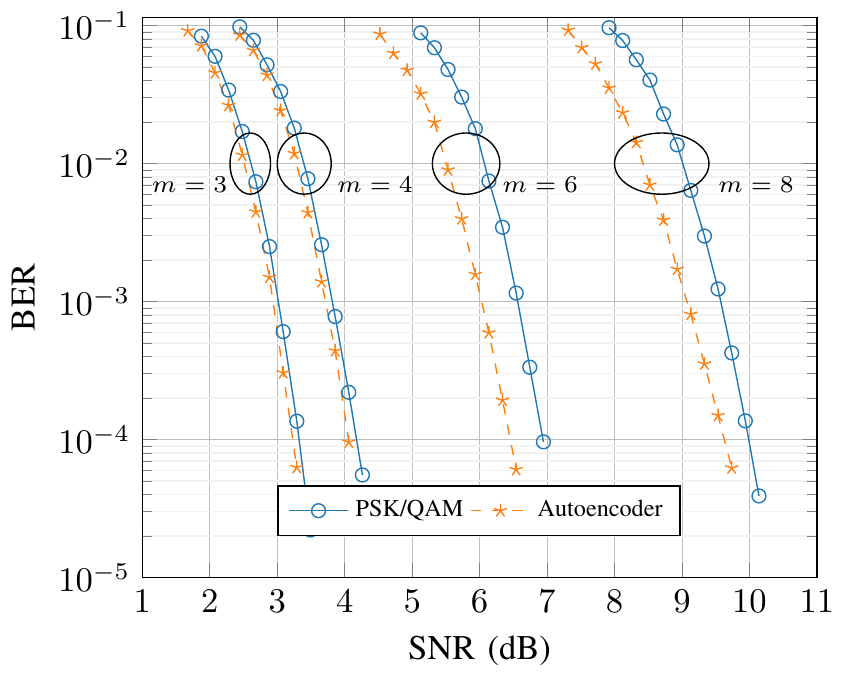}
	\caption{BER achieved by PSK/QAM constellations and the bit-wise autoencoder for $m=3,~4,~6$ and $8$. The baseline is PSK for $m = 3$, and QAM otherwise.\label{fig:ber-nid}}
\end{figure}

%% file: tex/it_ae_framework.tex
\begin{figure*}
 	\centering
  	\begin{subfigure}[b]{0.32\linewidth}
		\centering
		\tikzsetnextfilename{idd}
		\resizebox{1.0\linewidth}{!}{\includegraphics{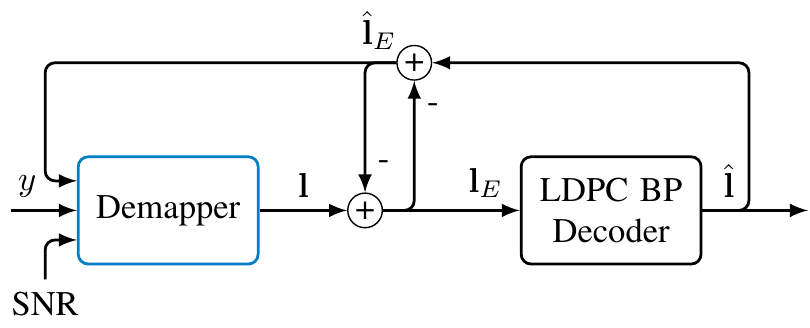}}
		\caption{{Iterative demapping and decoding} \label{fig:idd}}
	\end{subfigure} \quad
 	\begin{subfigure}[b]{0.63\linewidth}
 		\centering
 		\tikzsetnextfilename{idd_ae_setup}
		\resizebox{1.0\linewidth}{!}{\includegraphics{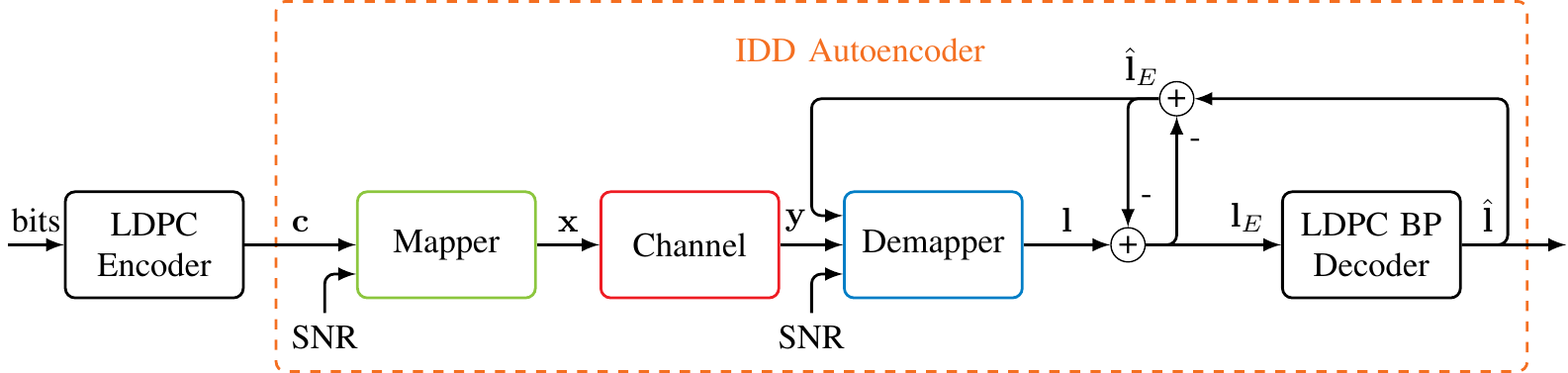}}
		\caption{IDD autoencoder setup\label{fig:setup-id}}
	\end{subfigure}
	\caption{Bit-wise autoencoder and iterative receiver structure. Each \ac{LDPC} codeword $\cv$ is mapped into $s$ bit vectors $\bv^{(i)}$, i.e., $s$ autoencoder transmissions. \label{fig:IDD-AE}}
\end{figure*}

In this section, the bit-wise autoencoder is extended to \ac{IDD} receivers~\cite{775793}.
The key idea of \ac{IDD} is to apply the Turbo Principle to soft demapping together with channel decoding, as illustrated in Fig.~\ref{fig:idd}.
With \ac{IDD}, the demapper takes as input the received samples $y$ together with prior knowledge on the bits denoted by $\hat{\lv}_E$, and generates soft decisions $\lv$.
$\hat{\lv}_E$ is subtracted from $\lv$, which results in the so-called \emph{extrinsic information} $\lv_E$ that is fed to the decoder.
The decoder then computes improved \ac{LLR}s based on the error-correcting code. These are denoted by $\hat{\lv}$.
The prior knowledge fed to the demapper is the extrinsic information $\hat{\lv}_E$ generated by the decoder, defined as $\hat{\lv}_E = \hat{\lv} - \lv_E$.
At the first iteration, $\hat{\lv}_E$ is set to the null vector, as no prior information is assumed to be available on the bits.
\ac{IDD} compensates for the information loss caused by \ac{BMD}, as the use of additional prior knowledge about the other bits result in a more reliable bit estimate.
Demapping with no prior knowledge gives separate estimates per bit, with no other prior knowledge than the channel output.
However, the extrinsic information feedback from the channel decoder allows to refine the demapper's estimate, which then refines the decoder estimate and vice versa.

The remainder of this section shows how the bit-wise autoencoder can be extended to \ac{IDD}.
This enables training of the end-to-end system including channel decoding with \ac{IDD} for any channel model.

\subsection{Extending the autoencoder with IDD}\label{sec:idd_ae}

\begin{algorithm}[t]
\SetAlgoLined
\SetKwInOut{Input}{Input}
\SetKwInOut{Output}{Output}
\SetKwBlock{Repeat}{repeat}{}
\SetKwFor{RepTimes}{For}{do}{end}
\DontPrintSemicolon
\Input{$y_1,\dots,y_s$}
\Output{$\hat{\mathbf{l}} = \LSB \hat{l}_1,\dots,\hat{l}_n \RSB\tp$}
\BlankLine
\tcc{Initialization}
$\hat{\lv}_E^{(j)[0]} \gets \zerov,~\forall j = 1,\dots,s$ \label{lst:init-prior}\\
$\mu_{c,v}^{[0]} \gets 0,~\forall (c,v) \in \Ec \label{lst:init-mcv}$
\BlankLine
\tcc{Iterative demapping \& decoding}
\RepTimes{$i=1,\dots,I$}{
\tcc{Demapping}
$\lv^{(j)[i]} \gets g_{\boldsymbol{\theta_D}} \LB y_i,\hat{\lv}_E^{(j)[i-1]} \RB,~\forall j = 1,\dots,s$ \label{lst:demap}\\
$\lv_E^{(j)[i]} \gets \lv^{(j)[i]} - \hat{\lv}_E^{(j)[i]},~\forall j = 1,\dots,s$ \label{lst:demap-ext}\\
$\lv_E^{[i]} \gets \LSB \lv_E^{(1)[i]\mathsf{T}},\dots,\lv_E^{(s)[i]\mathsf{T}} \RSB \tp$ \label{lst:demap-ext2}\\
\tcc{Decoding}
$\mu_{v,c}^{[i]} \gets l_{E,c}^{[i]} + \sum_{c' \in \Cc_v \setminus \{c\}} \mu_{c',v}^{[i-1]},~\forall (v,c) \in \Ec$ \label{lst:bp-start}\\
$\mu_{c,v}^{[i]} \gets 2\tanh^{-1}\LB \prod_{v' \in \Vc_c \setminus \{v\}} \tanh \LB \frac{\mu_{v',c}^{[i]}}{2} \RB \RB,~\forall (c,v) \in \Ec$\\
$\hat{l}_j^{[i]} \gets \hat{l}_{E,j}^{[i]} + \sum_{c' \in \Cc_j} \mu_{c,j}^{[i]},~\forall j = 1,\dots,n \label{lst:bp-end}$\\
$\hat{\lv}_E^{[i]} \gets \hat{\lv}^{[i]} - \lv_E^{[i]}$ \label{lst:decod-ext}\\
}
\Return $\hat{\lv} = \LSB \hat{l}_1^{[I]},\dots,\hat{l}_n^{[I]} \RSB\tp$
\caption{Iterative demapping and decoding\label{alg:id}}
\end{algorithm}

It is assumed that LDPC codes are used together with \ac{BP} decoding.
However, the proposed approach can be generalized to other coding schemes and decoding algorithms.
The autoencoder with \ac{IDD} receiver is shown in Fig.~\ref{fig:setup-id}, where $\lv_E^{(j)}$ and $\hat{\lv}_E^{(j)}$ denote the extrinsic information \cite{hagenauer2002turbo} corresponding to the $j^{th}$ complex baseband symbol, generated by the demapper and decoder, respectively.
The internal structure of the demapper remains the same as in Fig.~\ref{fig:demod-archi}. Only the number of inputs, and therefore the number of weights in the first layer, is increased.

The operations performed by the \ac{IDD} receiver are shown in Algorithm~\ref{alg:id}.
In this algorithm, $I$ denotes the number of iterations performed by the receiver, and $\Ec$ denotes the set of edges of the Tanner graph associated to the considered code.
$\Cc_v$($\Vc_c$) denote the neighboring check (variable) nodes of the variable (check) node $v$ ($c$). Some short background on \ac{LDPC} codes and terminology will be provided in Section~\ref{sec:cd}.
The superscript $[i]$ is used to refer to the $i^{th}$ iteration.
At each iteration, and for each received sample $y_j,~j=1,\dots,s$, the demapper $g_{\boldsymbol{\theta_D}}$ computes \ac{LLR}s on the bits, denoted by $\lv^{(j)}$, and using the previously computed extrinsic information generated by the decoder $\hat{\lv}_E^{(j)}$ (line~\ref{lst:demap}), initially set to zero (line~\ref{lst:init-prior}).
The extrinsic information generated by the demapper $\hat{\lv}_E^{(j)}$ for each received sample is then computed (line~\ref{lst:demap-ext}).
The extrinsic information generated by the demapper for the entire codeword is constructed by concatenating the ones of individual symbols (line~\ref{lst:demap-ext2}).
Next, one iteration of \ac{BP} is performed from the demapper extrinsic information (lines~\ref{lst:bp-start} to~\ref{lst:bp-end}).
Note that this requires knowledge of the values of messages from check nodes to variable nodes of the previous iteration, which are initially set to 0 (line~\ref{lst:init-mcv}).
Finally, the extrinsic information generated by the decoder is computed (line~\ref{lst:decod-ext}), which corresponds to the prior knowledge fed to the demapper at the next iteration.

\begin{figure*}
	\centering
	\tikzsetnextfilename{idd_unfolded}
	\resizebox{1.0\linewidth}{!}{\includegraphics{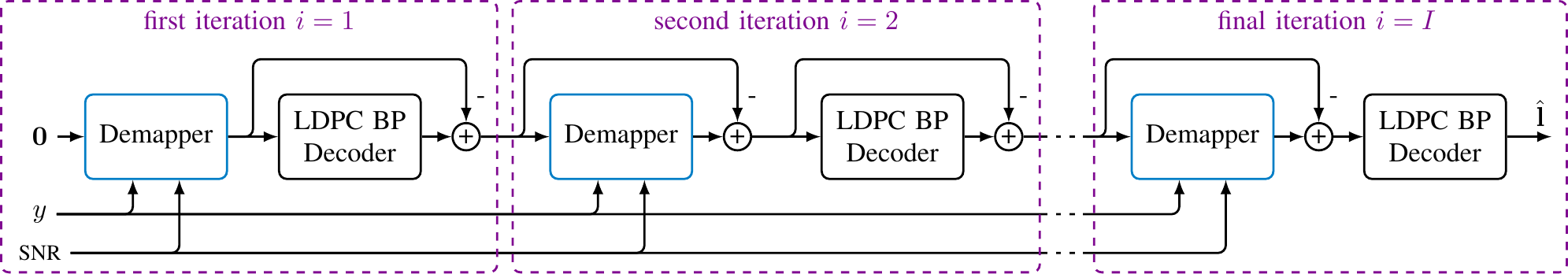}}
	\caption{Computational graph of the unfolded \ac{IDD} receiver.\label{fig:idd-unfold}}
\end{figure*}

As done in~\cite{7852251} with conventional \ac{BP}, the iterative algorithm depicted in~Algorithm~\ref{alg:id} is unfolded.
This is illustrated in Fig.~\ref{fig:idd-unfold}, where the first two iterations and the final iteration are shown, while intermediate iterations are straightforward and are therefore indicated by dashed lines.
The so obtained end-to-end system forms a feedforward \ac{NN}, which contains only differentiable operations.
Therefore, it can be trained in an end-to-end manner with usual backpropagation.
By doing so, the constellation, labeling, and demapper are jointly optimized for the considered channel and number of iterations. Although Algorithm~\ref{alg:id} uses the same trainable demapper $g_{\boldsymbol{\theta_D}}$ in each iteration, one could also consider a different demapper in each iteration. Additionally, one could enrich the BP equations with trainable weights, as done in \cite{7852251}. Also lower-complexity BP variants, such as min-sum decoding, can be used as long as they are differentiable.

Compared to the bit-wise autoencoder introduced in the previous section, which operates on individual complex basedband symbols carrying $m$ bits each, the autoencoder extended to \ac{IDD} operates on entire codewords of $n$ bits.
The loss (\ref{eq:loss-nid}) can be extended to such a system by
\begin{equation} \label{eq:loss-id-bad}
	\sum_{j=1}^n H \LB p_{\boldsymbol{\theta_M}}(c_j|\yv),\widetilde{p}_{\boldsymbol{\theta_D}}(c_j|\yv) \RB
\end{equation}
where summation is performed over the bits forming a codeword, and $\yv = \LSB y_1,\dots,y_s \RSB\tp$.
The value of a single bit $c_j$ depends on multiple baseband samples as channel decoding is considered as part of the autoencoder.
It was experimentally observed that training the end-to-end system by minimizing (\ref{eq:loss-id-bad}) leads to poor performance.
Therefore, the following loss is used, which is obtained by summing the binary \ac{CE} computed at the output of the demapper over all iterations:
\begin{align}
	\Jc(\boldsymbol{\theta_M}, \boldsymbol{\theta_D}) &\coloneqq \sum_{i=1}^I \sum_{j=1}^n H \LB p_{\boldsymbol{\theta_M}}(c_j|\yv),\widetilde{p}_{\boldsymbol{\theta_D}}^{[i]}(c_j|\yv) \RB \label{eq:loss-id}\\
		&= \sum_{i=1}^I \sum_{j=1}^n \Expect{\yv,c_j}{-\log{\widetilde{p}_{\boldsymbol{\theta_D}}^{[i]}(c_j|\yv)}} \label{eq:j_loss_2}
\end{align}
where $\widetilde{p}_{\boldsymbol{\theta_D}}^{[i]}(c_j|\yv)$ is the estimated posterior distribution at the $i^{th}$ iteration obtained by applying the sigmoid function to the logits generated by the demapper $\lv^{[i]} = \LSB \lv^{(1)[i]\mathsf{T}},\dots,\lv^{(s)[i]\mathsf{T}}\ \RSB\tp$.

\subsection{Simulation setup and results}

The iterative bit-wise autoencoder with \ac{IDD} was evaluated considering an \ac{AWGN} channel and
the same channel coding scheme as in Section~\ref{sec:nid-setup}.
Loop unrolling was applied to train through the differentiable \ac{IDD} receiver as shown in Fig.~\ref{fig:idd-unfold}. 
The number of iterations was set to $I = 40$, and the same set of weights was used in the demapper for all iterations. 
A benefit of this is that the number of weights does not increase with the number of iterations.
As in~~\ref{sec:nid-setup}, the Adam optimization algorithm \cite{kingma2014} was used and the learning rate was progressively decreased from $10^{-3}$ to $10^{-5}$ (see Tab~\ref{tab:training_param}).
The batch-size was set to 500 codewords.

\input{figs/ber_id.tex}

Fig.~\ref{fig:ber-id} shows the \ac{BER} achieved by \ac{PSK}/\ac{QAM} and the bit-wise autoencoder with and without \ac{IDD}.
Unsurprisingly, \ac{IDD} improves the performance of both the baseline and the autoencoder.
Moreover, one can see that the autoencoder achieves significant gains over standard modulations when considering \ac{IDD}.
Interestingly, the autoencoder without \ac{IDD} achieves \acp{BER} lower than the ones of the baseline with \ac{IDD} for high modulation orders (64 ($m=6$) and 256 ($m=8$)).

%% file: figs/ber_id.tex
\newlength{\rxx}
\newlength{\ryy}

\begin{figure}
		\tikzsetnextfilename{ber-id}
\includegraphics{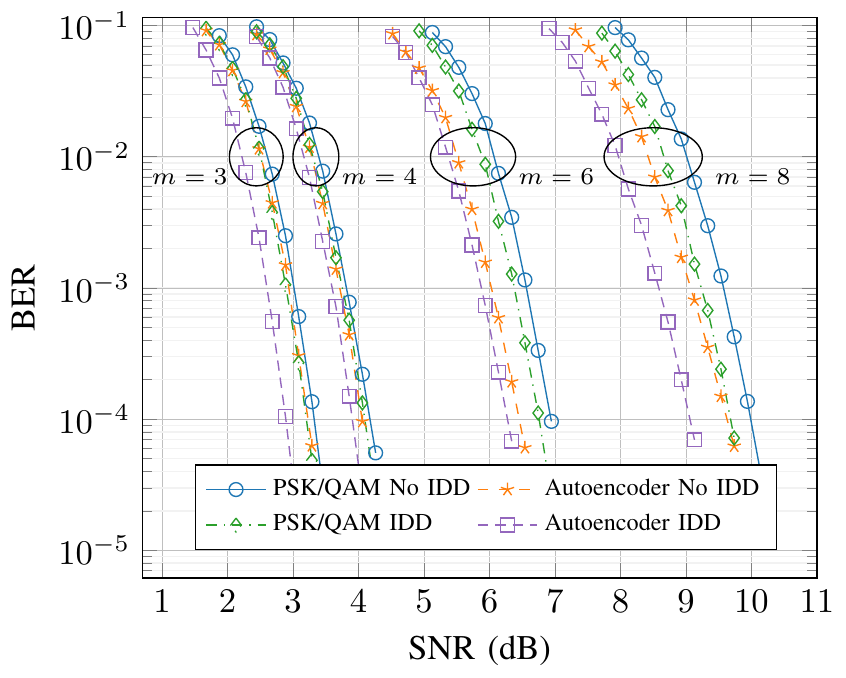}
	\caption{\ac{BER} achieved by PSK/QAM constellations with and without IDD and the bit-wise autoencoder with IDD for $m=3,~4,~6$, and $8$. The baseline is PSK for $m = 3$, and QAM otherwise.\label{fig:ber-id}}
\end{figure}

%% file: tex/code_opt.tex
\begin{figure}
  	\begin{subfigure}{\linewidth}
  		\centering
		\includegraphics[scale=0.40]{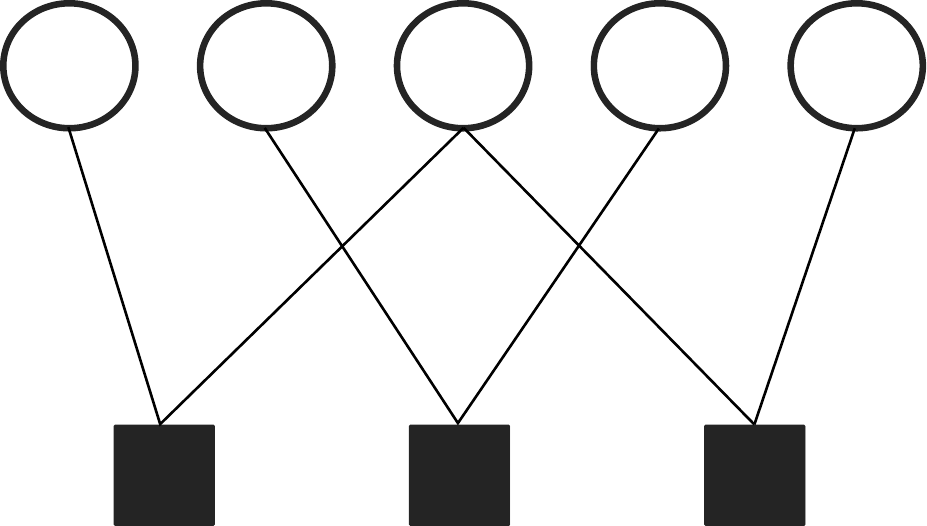}
		\caption{A Tanner graph: Circles depict \acp{VN} and squares \acp{CN}, respectively\label{fig:ldpc-ex-g}}\vspace*{0.4cm}
	\end{subfigure}
 	\begin{subfigure}{\linewidth}
		\centering
        $\begin{pmatrix}
            1 & 0 & 1 & 0 & 0\\
            0 & 1 & 0 & 1 & 0\\
            0 & 0 & 1 & 0 & 1\\
        \end{pmatrix}$
 		\caption{Parity-check matrix equivalent to the Tanner graph in (a)\label{fig:ldpc-ex-m}}
	\end{subfigure}
	\caption{A Tanner graph and its corresponding parity-check matrix\label{fig:ldpc-ex}}
\end{figure}

The previous sections considered an \ac{LDPC} code from the 802.11n standard~\cite{5307322}.
This section investigates how further gains can be achieved by optimization of the code after the system is trained. As we aim to communicate over arbitrary channels, off-the-shelf codes are not necessarily optimal and, as such, may limit the achievable overall performance.
An \ac{LDPC} code can be described by a Tanner graph which is composed of two types of nodes, \acp{VN} and \acp{CN}, representing bits and parity-check constraints, respectively.
An exemplary Tanner graph and its equivalent parity-check matrix are shown in Fig.~\ref{fig:ldpc-ex}.
The node degree corresponds to the number of edges that are connected to a node. In this example, all \acp{CN} have a degree of two, while the \acp{VN} have a degree of either one or two. The degree distribution considering all nodes is referred to as degree profile.

The conventional \ac{LDPC} code design method (e.g., \cite{richardson_urbanke_2008},\cite{1291808}) is applied in this section to materialize the improved \ac{IDD} performance.
This approach is based on the construction of ensembles of codes whose average behavior approaches the fundamental Shannon limit \cite{chung2001design}.
An ensemble is characterized by its degree profile, i.e., codes from an ensemble share the same \ac{VN} and \ac{CN} degree distributions. Hereby, the optimization of ensembles is motivated by the concentration-around-ensemble-average property~\cite{910577}, which states that, as the length of codewords increases, codes from an ensemble tend to behave close to the ensemble average. The whole task of the code optimization is to find a degree distribution that allows successful iterative decoding while maximizing the rate of the code.

As usually done, we introduce the polynomials $\rho(z) = \sum_{j\geq1}\rho_j z^{j-1}$ and $\lambda(z) = \sum_{i\geq1}\lambda_i z^{i-1}$, where $\rho_j$ and $\lambda_i$ denote the fractions of edges that connect to a \ac{CN} of degree $j$ and a \ac{VN} of degree $i$, respectively. Also describing the same code ensemble as the previously introduced \emph{node degree distribution}, the edge perspective simplifies the later analysis.
Note that $\forall j, \rho_j \geq 0,\forall i, \lambda_i \geq 0$, and $\sum_{j\geq1}\rho_j = \sum_{i\geq1}\lambda_i = 1$. Further, the average \ac{VN} degree $v_\text{avg}$ can be calculated as $v_\text{avg}=\nicefrac{1}{\int_{0}^{1}\lambda(x)\text{d}x}=\nicefrac{1}{\sum_{i\geq2}\frac{\lambda_i}{i}}$ and the average \ac{CN} degree $c_\text{avg}$ can be calculated as $c_\text{avg}=\nicefrac{1}{\sum_{j\geq2}\frac{\rho_j}{j}}$, respectively (see \cite{910577}).
A well-performing ensemble is constructed by optimizing $\rho$ and $\lambda$, such that the design rate
\begin{equation}
	\label{eq:drate}
	r = 1 - \frac{v_\text{avg}}{c_\text{avg}} = 1 - \frac{\sum_{j\geq2}\frac{\rho_j}{j}}{\sum_{i\geq2}\frac{\lambda_i}{i}}
\end{equation}
is maximized, while \ac{BP} decoding still converges to a zero average bit error probability.
This is achieved using the \ac{EXIT} chart approach~\cite{1291808}.
With this approach, it is assumed that the \acp{LLR} fed to the demapper as prior knowledge, denoted by $\hat{\lv}_{E}$, as well as those fed to the decoder are Gaussian distributed.
This is a common and valid assumption considering a large enough number of iterations and sufficiently long codewords. 

\subsection{EXIT analysis} \label{sec:exit-primer}

Recall that for iterative receiver implementations the extrinsic information denotes the new information gained by a receiver component (e.g., channel decoder, due to exploiting redundancy of the code) using \emph{a priori} information and, if available (e.g. for demapper, detection front-ends) the channel observations.
For the concept of iterative receivers (cf. \emph{Turbo} principle \cite{hagenauer2002turbo}), only the extrinsic information per component should be forwarded.
However, if we directly maximize the output \ac{MI} of the autoencoder as in Sec.~\ref{sec:idd_ae}, the resulting \ac{MI} violates this condition as the a priori information will be part of the demapper's output. Thus, we explicitly subtract the a priori knowledge in the \ac{IDD} setup in Fig.~\ref{fig:IDD-AE}.
A theoretical derivation of the demapper's extrinsic information becomes mathematically intractable due to the trainable weights in the receiver and we have to rely on Monte-Carlo sampling of the mutual information. This procedure will be described in the following and we refer to \cite{hagenauer2002turbo,ten2001convergence} for details.

We are interested in the \ac{MI} between the random variable associated with the input bit sequence $B$ and another random variable of their corresponding \ac{LLR} representations $L_E$ at the demapper's output. 
Unfortunately, this requires knowledge of the a posterior distribution $p(L_E|B=b)$ (see \cite{cover2012elements} for details) which could be approximated by measuring the corresponding histograms, or, more simply, by the following approximation from \cite{hagenauer2002turbo} 
$$I(B; L_E) = \left(1 - \frac{1}{n} \sum_{i=1}^n \left[ \operatorname{log_2} \left(1 + e^{-l_{E,i}\left(-2b_i+1\right)} \right) \right] \right)$$
where $n$ denotes a large enough amount of samples.

For the \ac{EXIT}-chart analysis, it is convenient to assume the transmission of the all-zero codeword. However, this is only valid for symmetric channels which is obviously not the case for autoencoder-based transmission, i.e., in general $p(Y=y|c=0) \neq p(Y=-y|c=1)$. We ensure symmetric channels (during code design) by introducing a pseudo random scrambling sequence known to both transmitter and receiver, e.g., see \cite{hou2003capacity}. This essentially means that random bits $\cv$ are transmitted, but at the receiver (i.e., the demapper output) the signs of the corresponding \acp{LLR} $\lv_E$ are flipped if $c_i=1$ (as if the all-zero codeword was transmitted). Note that, for readability, this detail is omitted in the following and the code optimization is explained as if the all-zero codeword was transmitted.

To make the mutual information accessible for Monte-Carlo simulations, the relation between mutual information and \acp{LLR} is required. Under the all-zero codeword assumption, the mutual information shared between the transmitted bits and Gaussian distributed \acp{LLR} with the mean value $\mu$ (and variance $2\mu$ \cite{1291808}) can be quantified by 
\begin{equation}
	J(\mu) = 1-\int_{-\infty}^{\infty} \frac{e^{-(\tau-\mu)^2/(4\mu)}}{\sqrt{4\pi\mu}} \operatorname{log}(1+e^{-\tau})d\tau
\end{equation}
thus, leading to $I(B_j;L_j)=J\left(\mathbb{E}_{l_j} \left[ l_j \right]\right)$.
Note that $J(.)$ takes values in the interval $(0,1)$, and a numerical approximation is given in~\cite{1291808}. 

\ac{EXIT} chart-based \emph{degree profile matching} requires knowledge of the mutual information between the output \acp{LLR} of the demapper and the transmitted bits\footnote{We assume a random interleaver between demapper and decoder.} which can be simulated via Monte-Carlo methods as shown later. In the case of \ac{IDD}, the demapper observes $y$ and additionally $\hat{\lv}_{E}$ as side information (so-called \emph{a priori knowledge} denoted by the mutual information $\tilde{I}_A$). However, for the code design phase (all-zero codeword) we replace $\hat{\lv}_{E}$ by Gaussian distributed \acp{LLR} $\tilde{\lv}_E$ such that the mean of these a priori \acp{LLR} becomes $\tilde{\mu} = J^{-1} (\tilde{I}_A)$ (see~\cite{1291808} for a numerical approximation), which, again, is much simpler to simulate.
The demapper's output is a vector with soft-estimates on all $m$ transmitted bits. However, in the \ac{BICM} case, these $m$ estimates are treated equally and the demapper's output can be described by the mean value of its output \acp{LLR} $\lv_E$.

\begin{figure}[t]
  	\centering
  	\tikzsetnextfilename{exit_idd}
	\includegraphics{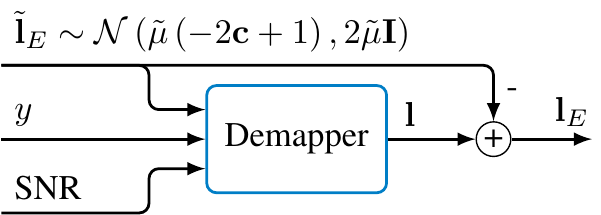}
	\caption{EXIT-analysis of the demapper assuming Gaussian distributed a priori \acp{LLR} $\tilde{\lv}_{E}$\label{fig:idd-exit}}
\end{figure}

The demapper's \ac{EXIT} characteristic is denoted by $T(.)$ and describes the input/output relation of the \emph{extrinsic} mutual information of the iterative demapper~\cite{1291808}, i.e., for a given a priori mutual information, how much (additional) extrinsic information the demapper can contribute by observing the channel output.
This \ac{EXIT} characteristic is defined\footnote{For readability, we define $T(\mathrm{SNR},\mu)$ as the input/output relation of the mean of the \acp{LLR}. However, this directly relates to the mutual information via the $J(.)$-function.} as
\begin{multline}
	T\LB \mathrm{SNR}, \tilde{\mu} = J^{-1}\LB \tilde{I}_A \RB \RB\\
	= J^{-1} \LB \frac{1}{m} \sum_{i=1}^m I \LB B_i;L_{E,i} | \tilde{L}_{E,i} \RB \RB
\end{multline}
where $L_{E,i}$ denotes the demapper's extrinsic soft estimate on the $i$\textit{th} bit and $\tilde{L}_{E,i}$ is chosen such that $\tilde{I}_A = I(B_i; \tilde{L}_{E,i})$.
This can be achieved by sampling such that $\tilde{\lv}_E \sim \Nc \LB \tilde{\mu} \LB -2\cv+1 \RB, 2\tilde{\mu}\Id \RB$ as shown in Fig.~\ref{fig:idd-exit}.
Note, that from code design perspective, this transfer function $T(.)$ characterizes the \emph{channel front-end}, i.e., $T(.)$ depends on the trained systems so that the optimum code can only be defined after training. As such, the training of the autoencoder impacts the optimal degree profile.

A closed form solution for $T(.)$ is difficult to find, however, for a fixed \ac{SNR} and given $\tilde{I}_A$, Monte Carlo-based approaches are possible based on measured \acp{LLR} $l_{E,i}$.
We follow \cite{hagenauer2002turbo} for a numerical approximation of the mutual information for simulated \acp{LLR} $l_{E,i}$ leading to
\begin{multline}
	T \LB \mathrm{SNR},J^{-1} \LB \tilde{I}_A \RB \RB\\
	\approx J^{-1} \left(1 - \frac{1}{n} \sum_{i=1}^n \left[ \operatorname{log_2} \left(1 + e^{-l_{E,i}\left(-2c_i+1\right)} \right) \right] \right) \label{eq:mi_estimation}
\end{multline}
where $c_i$ denotes the codeword bit corresponding to the LLR $l_{E,i}$ and $n$ is sufficiently large. For the $J(.)$ function, we use the numerical approximation from~\cite{1291808}.
This turns out to be more precise in non-Gaussian scenarios than the naive mean-based estimation.

\subsection{Code design method} \label{sec:code-design}

Following the conventional method (e.g., \cite{richardson_urbanke_2008},\cite{1291808}), optimization of the code ensemble can be reduced to a linear optimization problem by fixing one of the two polynomials defining its degree profile and optimizing the respective other (cf. \cite{ref_id:webdemo}).
We choose a check-concentrated degree distribution $\rho_c = 1$ for some integer $c$, i.e., a CN-regular code of degree $c$ (e.g., $c=2$ in Fig.~\ref{fig:ldpc-ex-g}).
Therefore, maximizing the design rate~(\ref{eq:drate}) is equivalent to maximizing $\sum_{i=2}^{d}\frac{\lambda_i}{i}$, where $d$ is the maximum \ac{VN} degree.
Optimization of the \ac{VN} degree distribution is performed by solving the following optimization problem:
\begin{subequations}
\label{eq:ldpc-design-opt}
\begin{alignat}{2}
&\underset{\lambda_2,\dots,\lambda_d}{\operatorname{maximize}}	\quad	&&\sum_{i=2}^{d}\frac{\lambda_i}{i} \nonumber\\
&\text{subject to} 											  	\quad   &&\lambda_i \geq 0,~\forall i \in \{2,\dots,d\} \nonumber \\
&																\quad 	&&\sum_{i=2}^{d} \lambda_i = 1 \nonumber \\
&																\quad   &&\sum_{i=2}^{d} \lambda_i J \LB T \LB \text{SNR}, J^{-1} \LB h \RB \RB  + (i-1) J^{-1} \LB h \RB \RB \nonumber \\
&																\quad 	&& > 1 - J \LB \frac{1}{c-1}J^{-1} \LB 1 - h \RB \RB,~\forall h \in (0,1) \label{eq:conv-const} \\
&																\quad 	&& \lambda_2 \le \beta \label{eq:l2-const}
\end{alignat}
\end{subequations}

The constraint (\ref{eq:conv-const}) guarantees convergence to a zero average bit error probability (see, e.g.,~\cite{richardson_urbanke_2008} for more details).
The constrained (\ref{eq:l2-const}) provides a stability condition on $\lambda_2$ to ensure convergence close to the point $(1,1)$ in the EXIT chart as intersections between the VN and CN EXIT curves are numerically hard to track in this region. However, these intersections may translate into a non-negligible error-floor or \ac{BER} performance degradation. For the \ac{AWGN} case, it can be shown that $\beta \le \frac{ \exp{ \LB \frac{1}{N_0}\RB } } {c-1}$. However, $\beta$ also allows to generally control the amount of degree-2 \acp{VN} to simplify the later code construction phase as these degree-2 nodes require a careful (and complex) placement to avoid high error-floors.

\begin{algorithm}[tb]
\SetAlgoLined
\SetKwInOut{Input}{Input}
\SetKwInOut{Output}{Output}
\SetKwBlock{Repeat}{repeat}{}
\SetKwFor{RepTimes}{For}{do}{end}
\DontPrintSemicolon
\Input{Initial SNR, $r^*, \Delta_{\text{SNR}}, \epsilon, c, D$}
\Output{$\lambda^*_2,\dots,\lambda^*_v$}
\Repeat{
	Sample: $B$ codewords $\cv^{(1)},\dots,\cv^{(B)}$ \label{lst:k-est-b}\\
	Sample: corresponding observations $\yv^{(1)},\dots,\yv^{(B)}$\\
	\RepTimes{$i=1,\dots,D$}{
		$\tilde{\mu} \gets J^{-1}(\frac{i}{D})$\\
		Sample: $\tilde{\lv}_E^{(1)},\dots,\tilde{\lv}_E^{(B)},~\tilde{\lv}_E^{(j)} \sim \Nc \LB \tilde{\mu} \LB -2\cv^{(j)}+1 \RB, 2\tilde{\mu}\Id \RB$\\
		Estimate $T \LB \text{SNR}, J^{-1}\LB \frac{i}{D} \RB \RB \approx J^{-1} \left(1 - \frac{1}{Bn} \sum_{k=1}^B \sum_{j=1}^n \right.$ \\ \hspace*{2cm} $\left.	 \left[ \operatorname{log_2} \left(1 + e^{-l_{E,j}^{(k)}\cdot\left(-2c_j^{(k)}+1\right)} \right) \right] \right)$ \label{lst:k-est-e}\\
	}
	Set $\lambda_2,\dots,\lambda_v$ by solving~(\ref{eq:ldpc-design-opt}) \label{lst:solve}\\
	$r \gets 1 - \frac{1}{c\sum_{j=2}^v \frac{\lambda_j}{j}}$ \label{lst:rate}\\
	\uIf{$r > r^*$}{ \label{lst:snr-dec-1}
		$\text{SNR} \gets \text{SNR} - \Delta_{\text{SNR}}$ \label{lst:snr-dec-2}\\
		$\lambda^*_2,\dots,\lambda^*_v \gets \lambda_2,\dots,\lambda_v$ \label{lst:snr-dec-3}
	}
	\uElse{ \label{lst:snr-inc-1}
		$\text{SNR} \gets \text{SNR} + \Delta_{\text{SNR}}$ \label{lst:snr-inc-2}
	}
	$\Delta_{\text{SNR}} \gets \frac{\Delta_{\text{SNR}}}{2}$\\
	\uIf{ $\Delta_{\text{SNR}} < \epsilon$ }{
		\Return $\lambda^*_2,\dots,\lambda^*_v$ 
	}
}
\caption{Code design algorithm}
\label{alg:code-design}
\end{algorithm}

In practice, the optimization problem~(\ref{eq:ldpc-design-opt}) is made tractable by uniformly discretizing the interval $(0,1)$ with $D$ values. Thus,~(\ref{eq:conv-const}) is replaced by the set of constraints
\begin{multline} \label{eq:quant-conv-const}
\sum_{i=2}^{d} \lambda_i J \LB T \LB \text{SNR}, J^{-1} \LB \frac{j}{D} \RB \RB  + (i-1) J^{-1} \LB \frac{j}{D} \RB \RB \\
	> 1 - J \LB \frac{1}{c-1}J^{-1} \LB 1 - \frac{j}{D} \RB \RB,~j = 0,\dots,D-1.
\end{multline}
Moreover, because $T(.)$ depends on the \ac{SNR},~(\ref{eq:ldpc-design-opt}) is specific to a given \ac{SNR} value.
However, one wants to find a code which achieves a desired target rate denoted by $r^*$ with the lowest possible \ac{SNR}.
To achieve this goal,~(\ref{eq:ldpc-design-opt}) is repeatedly solved until the \ac{SNR} cannot be decreased without achieving a rate lower than $r^*$, as shown in Algorithm~\ref{alg:code-design}.
In this algorithm, $T(.)$ is numerically estimated by the previously introduced Monte-Carlo approach (lines~\ref{lst:k-est-b} to~\ref{lst:k-est-e}) for a fixed \ac{SNR}.
Using this estimation, the optimization problem~(\ref{eq:ldpc-design-opt}), with~(\ref{eq:conv-const}) is replaced by~(\ref{eq:quant-conv-const}), is solved using a conventional linear solver (line~\ref{lst:solve}). The rate of the so-obtained solution $\lambda(z)$ is computed (line~\ref{lst:rate}). As~(\ref{eq:ldpc-design-opt}) aims at finding the degree profile $\lambda(z)$ that maximizes the rate for the given SNR and under the constraint that successful decoding is possible (constraint (\ref{eq:conv-const}) and (\ref{eq:l2-const})), this solution provides the largest possible rate for the given \ac{SNR}, $c$ and demapper $T(.)$.
If the rate is higher than the targeted rate $r^*$, the \ac{SNR} is decreased for the next iteration by a predefined amount $\Delta_{\text{SNR}}$ (lines~\ref{lst:snr-dec-1} to~\ref{lst:snr-dec-3}). Otherwise, it is increased (lines~\ref{lst:snr-inc-1} and~\ref{lst:snr-inc-2}). This describes a simple bisection search in a certain \ac{SNR} interval.
The optimization is repeated until $\Delta_{\text{SNR}}$ is below a predefined threshold $\epsilon$.\footnote{Note that the algorithm could be modified such that the \ac{SNR} is constant and the rate is maximized; depending on the exact setup, this can be beneficial, e.g., for transmissions where the \ac{SNR} is given.}
The same algorithm can be repeated for multiple values of $c$ until the best threshold is achieved.

Once the ensemble degree distribution is optimized, a single code from this ensemble is selected using conventional code construction methods (see, e.g.,~\cite{richardson_urbanke_2008}).

\subsection{Evaluation}

\begin{figure}
	\tikzsetnextfilename{exit_chart}
	\includegraphics{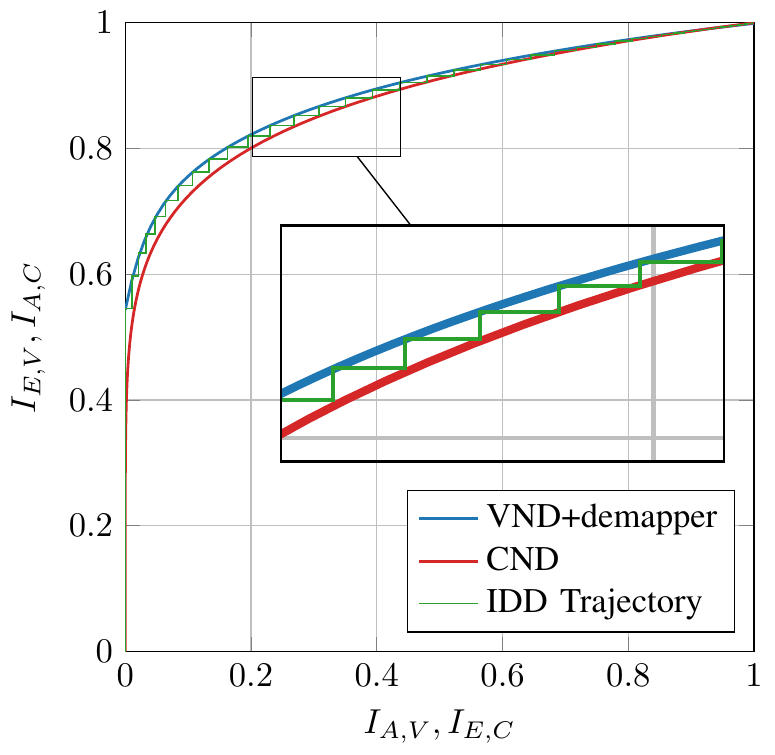}
	\caption{EXIT chart of an optimized LDPC code including the actual trajectory during decoding for the bit-wise autoencoder with $m=6$. The schemes use IDD with 80 iterations.\label{fig:exit}}
\end{figure}

\input{figs/ber_code_1944_id80.tex}

We use a \ac{PEG}-based code construction \cite{hu2005regular} to ensure that an optimized decoding threshold also translates into an improved \ac{BER} performance. %
The intuition behind \ac{PEG} is to maximize the \emph{girth}, i.e., to minimize short cycles in the graph and, thus, to avoid a performance degradation due to correlated messages during decoding.
The baseline uses the conventional 802.11n code of the same length of $n=1944$ bits. Further, we limit the amount of degree-2 VNs to be less than the number of CNs (e.g., in this case $\beta=0.26$). This helps to reduce error-floors at the price of a slightly decreased waterfall performance. However, in particular for short (to medium) code lengths, the explicit graph construction becomes a challenging task. 

Fig.~\ref{fig:exit} shows the corresponding EXIT-chart of the optimized code and the actual decoding trajectories during decoding. Note that $I_{A,V},I_{A,C}$ and $I_{E,V},I_{E,C}$ denote the a priori and extrinsic mutual information of the \ac{VN} and \ac{CN}, respectively. 
The \ac{VN} and \ac{CN} curves can be analytically described for a given ensemble (i.e., degree profile; see~\cite{1291808}) and describe the input/output characteristic of the extrinsic mutual information of the \ac{VN} decoder (+demapper) and the \ac{CN} decoder update, respectively. For a given input mutual information $I_A$, the curves trace how much new (i.e., extrinsic) mutual information $I_E$ can be gained with an \ac{VN} or \ac{CN} decoding update, respectively. If the \ac{VN} and \ac{CN} curves intersect, no successful decoding is possible as the iterative update of \acp{VN} and \acp{CN} does not generate sufficient extrinsic knowledge to progress convergence.
However, the plotted trajectory is the actual simulated performance during decoding. As can be seen, the iterative decoding trajectory step-wise tracks the predicted curves and, thus, follows its expected behavior. Further, it can be seen that the EXIT-curves are well-matched while maintaining an open tunnel during all decoding iterations, verifying the effectiveness of our code optimization.

\begin{figure*}
 	\centering
 	\begin{subfigure}[b]{0.45\linewidth}
 		\centering
 		\tikzsetnextfilename{rx_ae_training}
		\includegraphics{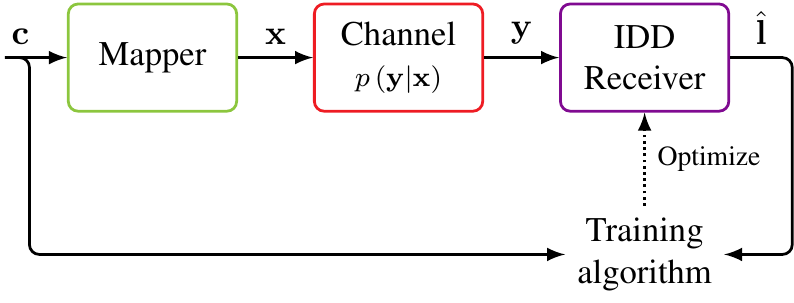}
		\caption{Training of the receiver\label{fig:rl_rx_train}}
	\end{subfigure} \hspace*{\fill}
 	\begin{subfigure}[b]{0.50\linewidth}
 		\centering
 		\tikzsetnextfilename{tx_ae_training}
		\includegraphics{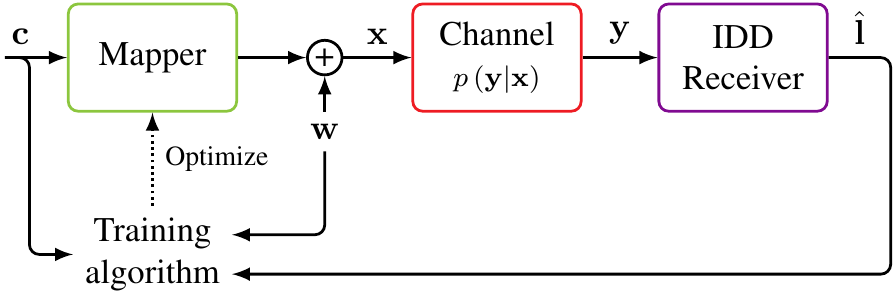}
		\caption{Training of the mapper without channel model\label{fig:rl_tx_train}}
	\end{subfigure} \qquad
	\caption{Training of the end-to-end system without a channel model is done by alternating between (a) conventional training of the receiver, and (b) training of the mapper using a method inspired by reinforcement learning.}
\end{figure*}

Fig.~\ref{fig:ber-code} shows the \ac{BER} performance of the optimized code with maximum \ac{VN} degree $v_{max} = 12$ for an \ac{AWGN} channel and for the 802.11n code of length $n=1944$ as a baseline. For both codes, both a QAM constellation and, as comparison, the bit-autoencoder for $m=6$ and $m=8$ are used. The code is optimized for the QAM and autoencoder individually to ensure the best performance of each system. 
Note that, for this canonical scenario, the baseline already provides a well-designed coding scheme. Interestingly, also the baseline benefits from the code optimization which is plausible as the 802.11n code was not specifically designed for \ac{IDD} receivers and, thus, can also be outperformed by a code carefully matched to the \ac{IDD} receiver. However, as proof-of-concept, the proposed approach outperforms the baseline by approx. 0.1 dB for $m=6$ and 0.2 dB for $m=8$, respectively, but larger gains are expected whenever $m$ increases, or when the channel becomes more sophisticated, as will become obvious in the next section.

%% file: figs/ber_code_1944_id80.tex
\begin{figure}
		\tikzsetnextfilename{ber-code}
\includegraphics{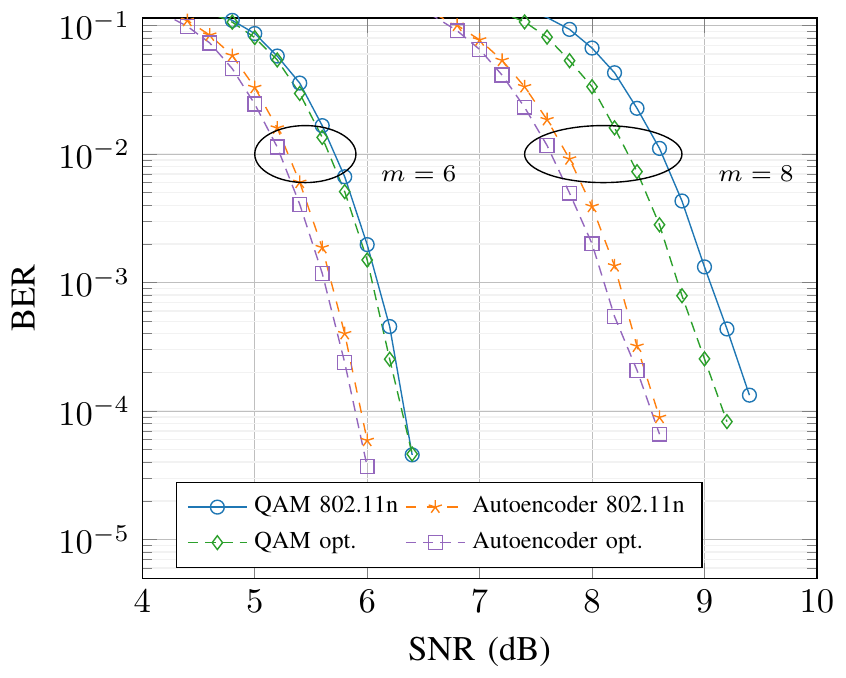}
	\caption{\ac{BER} achieved by baseline QAM constellations and the bit-wise autoencoder for $m=6$ and $m=8$ bits per symbol. We compare between 802.11n \ac{LDPC} codes \cite{5307322} and optimized \ac{LDPC} codes of the same length $n=1944$. All schemes use IDD with 80 iterations. \label{fig:ber-code}}
\end{figure}

%% file: tex/ota.tex
This section shows how the previously described system can be trained over an actual channel.
In previous sections, training of the autoencoder was performed considering an \ac{AWGN} channel model.
However, it was shown in~\cite{8214233} that training using a channel model leads to significant loss of performance once the system is deployed on the actual channel.
This loss of performance was attributed to the unavoidable mismatch between the channel model and the actual channel, and was observed despite important efforts put on the conception of an accurate channel model.
These results motivated the design of a training method for autoencoder-based communication systems in~\cite{8792076} that does not require a model of the channel.
Moreover, this method can be used for training over channels with non-differentiable components, such as quantization.

\subsection{Training over the actual channel}\label{sec:rl-training}

Following~\cite{8792076}, training of the end-to-end system is done in an alternating fashion in which each iteration is composed of two stages, one for the receiver and one for the mapper.
Training of the receiver, shown in Fig.\ref{fig:rl_rx_train}, does not require differentiation of the channel and is therefore conventional.

First, a batch of codewords is sampled, and each codeword $\cv = \LSB \bv^{(1)\tp},\dots,\bv^{(s)\tp} \RSB\tp$ is mapped onto a vector of channel symbols $\xv = f_{\boldsymbol{\theta_M}}(\cv) = \LSB f_{\boldsymbol{\theta_M}}(\bv^{(1)}), \dots, f_{\boldsymbol{\theta_M}}(\bv^{(s)}) \RSB\tp$.
Thus, $\xv$ is sent over the channel and the received symbols $\yv$ are fed to the \ac{IDD}-based receiver, generating \acp{LLR} $\hat{\lv}$.
Training is done using \ac{SGD} on the loss (\ref{eq:j_loss_2}).
At training, it is assumed that the receiver has knowledge of the transmitted codewords $\cv$ which can be achieved, e.g., through pseudorandom number generators initialized with the same seed on both ends.

Training of the transmitter is less straightforward as differentiation of the channel is not possible through conventional backpropagation.
Inspired by reinforcement learning, the method introduced in~\cite{8792076} relies on the addition of random perturbations $w$ to the output of the transmitter at training.
This approach enables approximation of the gradient of the loss function \ac{wrt} the transmitter parameters - despite the lack of a channel model.
From the viewpoint of reinforcement learning the addition of perturbations enables exploration of neighboring solutions that may lead to better performance.
Formally, the transmitter output is relaxed to a normal distribution by adding a zero-mean Gaussian perturbation $w \thicksim \CN(\zerov,\sigma_{w}^2)$
\begin{equation}
x = \sqrt{1-\sigma_{w}^2}f_{\boldsymbol{\theta_M}}(\bv) + w
\end{equation}
where scaling is done to ensure the conservation of the average energy, and $\sigma_{w}$ must be in the interval $(0,1)$.
Training of the transmitter is performed on the loss
\begin{equation}
	\widehat{\Jc}(\boldsymbol{\theta_M}) \triangleq \sum_{i=1}^I \sum_{j=1}^n \Expect{c_j,\wv,\yv}{-\log{\widetilde{p}_{\boldsymbol{\theta_D}}^{[i]}(c_j|\yv)}}
\end{equation}
instead of~(\ref{eq:j_loss_2}), where $\wv = [w_1,\dots,w_S]\tp$ is the vector of \ac{i.i.d.} perturbations added to the baseband symbols generated by the transmitter for a given codeword.
Note that, compared to~(\ref{eq:j_loss_2}), the expected value is also over the perturbations $\wv$, as $\yv \thicksim p(\yv|f_{\boldsymbol{\theta_M}}(\cv) + \wv)$.
The gradient of $\widehat{\Jc}$ is estimated by
\begin{multline} \label{eq:j_hat_est}
	\nabla_{\boldsymbol{\theta_M}}\widehat{\Jc}(\boldsymbol{\theta_M}) \approx \frac{1}{|\Vc|}\sum_{\cv \in \Vc}\sum_{j=1}^n\sum_{i=1}^I \Bigg( \frac{2\sqrt{1-\sigma_{w}^2}}{\sigma_{w}^2}\cdot\\
	\Big( c_j \log{\LB \widetilde{p}_{\boldsymbol{\thetav_D}}^{[i]} \LB c_j|\yv \RB\RB} + \LB 1 - c_j \RB \log{\LB 1 - \widetilde{p}_{\boldsymbol{\thetav_D}}^{[i]} \LB c_j|\yv \RB \RB} \Big)\cdot\\
	\nabla_{\boldsymbol{\theta_M}}f_{\boldsymbol{\theta_M}}(\cv) \wv \Bigg)
\end{multline}
where $\Vc$ is a finite set of samples of codewords, $\nabla_{\boldsymbol{\theta_M}}f_{\boldsymbol{\theta_M}}(\cv) = \LSB \nabla_{\boldsymbol{\theta_M}}f_{\boldsymbol{\theta_M}}(\bv^{(1)}),\dots,\nabla_{\boldsymbol{\theta_M}}f_{\boldsymbol{\theta_M}}(\bv^{(s)}) \RSB$, and the so called ``log-trick'' $\nabla_{\xv} \log{p(\yv|\xv)} = \frac{\nabla_{\xv}p(\yv|\xv)}{p(\yv|\xv)}$ was used.
It was shown in~\cite{8792076} that in cases where the channel is differentiable, $\nabla_{\boldsymbol{\theta_M}}\widehat{\Jc}(\boldsymbol{\theta_M})$ converges to the gradient of~(\ref{eq:j_loss_2}) $\nabla_{\boldsymbol{\theta_M}}\Jc(\boldsymbol{\theta_M})$ as $\sigma_w$ goes to zero.
Unfortunately, reducing $\sigma_w$ leads to higher variance of the gradient estimator~(\ref{eq:j_hat_est}), causing slow convergence.
Therefore, $\sigma_w$ controls a tradeoff between the accuracy of the loss function gradient approximation and the estimator variance.

Figure~\ref{fig:rl_tx_train} shows the training of the mapper.
Each codeword $\cv \in \Vc$ is mapped to a vector of baseband symbols $f_{\boldsymbol{\theta_M}}(\cv)$.
A perturbation vector $\wv \thicksim \Nc(\zerov, \sigma_w^2\Id)$ is added to $f_{\boldsymbol{\theta_M}}(\cv)$ to form the vector of symbols $\xv$ that is sent over the channel.
The received vector of symbols $\yv$ is demapped and decoded by the \ac{IDD}-based receiver, and the so-obtained \acp{LLR} $\hat{\lv}$ are sent to the mapper using a reliable link required only at training.
Finally, one step of optimization is performed with the loss gradient estimated by~(\ref{eq:j_hat_est}).
By alternating between a few steps of receiver training (Fig.~\ref{fig:rl_rx_train}) and a few steps of transmitter training (Fig.~\ref{fig:rl_tx_train}), optimization of the end-to-end system is carried out - despite the lack of a channel model.

\subsection{Experimental results}

\input{figs/ber_OTA_1944_id80.tex}

\input{figs/ber_OTA_1944_id80_map.tex}

A wireless communication system consisting of two \acp{USRP} B210 from Ettus Research with carrier-frequency of $2.35\, \text{GHz}$ and an effective bandwidth of $15.94\, \text{MHz}$ was trained in a static indoor office environment.
An \ac{OFDM}-based autoencoder with \ac{CP} of ratio $\nicefrac{1}{8}$ and 64 subcarriers (50 of which are used for data transmission) as introduced in~\cite{felix2018ofdm} was considered.
The channel was mostly frequency flat with varying \ac{SNR} between sub-carriers of at most $\approx 2$dB.
Although, the \ac{OFDM} framework was only used for synchronization and the \ac{CP} was not accessible to the \ac{NN}.
From the autoencoder's point of view it was operating as a single carrier system (within a sub-carrier of the \ac{OFDM} framework) and other sub-carriers and their channel conditions were unknown to it.
Linear \ac{MMSE} equalization of the received symbols was performed on a per-subcarrier-basis prior to \ac{NN}-based demapping.\footnote{In \cite{felix2018ofdm}, it has been shown that the autoencoder can learn \ac{MMSE} equalization, however, this requires more than one complex-valued channel use. For simplicity and a fair comparison with the \ac{QAM}-baseline, this case is not considered here, yet an extension is straightforward.}
We did not compensate for any other potential analog or digital impairments such as distortion, quantization or clipping.
To estimate the \ac{SNR}, we first calculate the \ac{EVM} between the originally sent symbols and the equalized received symbols.
We then calculate an average \ac{SNR} per sub-carrier, defined by the mean over the \ac{EVM}, and feed this \ac{SNR} estimation to the demapper of the corresponding sub-carrier.
To perform code optimization (see Section~\ref{sec:cd}), the \ac{EXIT} curves were measured based on data from actual over-the-air transmissions.
The \ac{NN} parameters were the same as in the previous sections, and $\sigma^2_{w}$ was progressively decreased from 0.05 to 0.01 at training.
The only exception is that we did not feed the current \ac{SNR} to the autoencoder encoder, as this would require a precise \ac{SNR} sounding prior to the actual transmission.
Instead for all \ac{OTA} results, we use a single fixed constellation trained within a certain \ac{SNR} range centered around the \ac{LDPC}'s waterfall region.

The obtained \ac{BER} curves are shown in Fig.~\ref{fig:ber-ota}.
Obviously, the autoencoder maintains its superiority when compared to the \ac{QAM} baselines and can even increase the performance gap to approximately 1.0 dB.
Code optimization provides further improvements of 0.2 dB and 0.4 dB for $m=6$ and $8$, respectively.
This can be intuitively explained by the fact that the 802.11n \ac{LDPC} code was optimized for an \ac{AWGN} channel, while for \ac{OTA} training, different constellations (and, thus, different demapper characteristics) may turn out to be optimal.

Next, we will focus on the reasons of why a trained system outperforms the \emph{classical} baseline over the actual channel.
Let us quickly recap the reasons for potential performance gains which are superimposed in Fig.~\ref{fig:ber-ota}:
\begin{enumerate}
\item Optimized constellations. This typically requires \ac{IDD} to materialize the full performance gains.
\item An optimal demapper for the actual channel. For the \ac{OTA} channel the \ac{AWGN}-\ac{MAP} demapper is a \emph{mismatched} demapper and, thus, sub-optimal for the actual channel. The AE can learn the underlying channel transition probabilities and, thereby, provides better \ac{LLR} estimates (i.e., closer to the actual true \emph{a posteriori probability}).
\item \ac{BER} gains due to the \emph{matched} \ac{LDPC} code, i.e., the \ac{IDD} receiver can recover more information from the received sequence. %
\end{enumerate}

We compare the \ac{BER} achieved by an explicit \ac{AWGN}-\ac{MAP} demapper for \ac{IDD} (see \cite{775793}), only optimal for the \ac{AWGN} channel, and the learned autoencoder constellation as shown in Fig~\ref{fig:ber-ota-map}.
Note that, this requires knowledge of the underlying channel statistics and only serves as an optimal baseline for the canonical \ac{AWGN} channel, i.e., it provides a lower bound of the achievable \ac{AWGN} performance. It can be seen that on the \ac{AWGN}, the autoencoder-based receiver virtually approaches the same \ac{BER} performance, and, thus, can be claimed to be optimal for the given \ac{AWGN} channel.
For the \ac{OTA} transmission, the learned demapper yields a gain of 0.6~dB compared to the \ac{MAP} demapper for the \ac{AWGN} channel period.\footnote{To account for alterations after deployment, it is possible to continuously finetune the autoencoder demapper by a simple supervised training step as shown in~\cite{schibisch2018online} The labels for such a re-training can be directly taken from the \ac{BP} decoder's output $\hat{\lv}$ which can be assumed to be (nearly) error-free for a typical operation point. Thereby, continuous receiver finetuning without any additional piloting overhead is possible.}.
This can be intuitively explained as even after equalization, the actual channel is not necessarily \ac{AWGN}, resulting in suboptimal performance for the \ac{MAP} demapper when an \ac{AWGN} channel is assumed.
The learned demapper was, on the other hand, optimized on the actual channel, leading to its superior \ac{BER} performance.

%% file: figs/ber_OTA_1944_id80.tex
\begin{figure}
		\tikzsetnextfilename{ber-ota}
\includegraphics{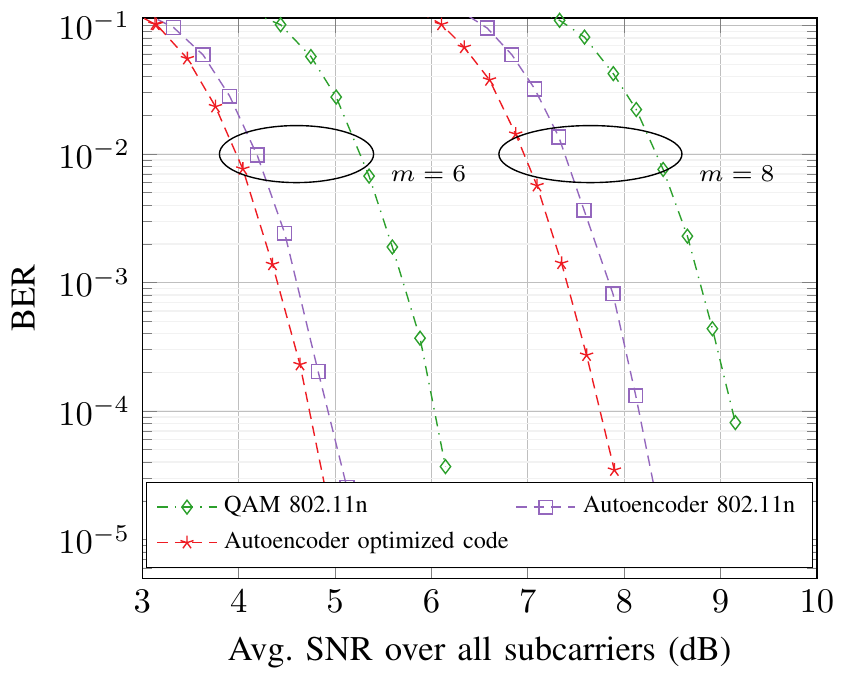}
	\caption{Actual \ac{OTA} \ac{BER} achieved by baseline QAM constellations and the bit-wise autoencoder for $m=6$ and $8$. The baseline uses the 802.11n code of same length $n=1944$ as the optimized (opt.) code (see Section \ref{sec:cd}). The SNR is averaged over all subcarriers of the OFDM system. All schemes use IDD with 80 iterations. \label{fig:ber-ota}}
\end{figure}

%% file: figs/ber_OTA_1944_id80_map.tex
\begin{figure}

\begin{subfigure}[t]{\columnwidth} \centering
		\tikzsetnextfilename{ber-ota-map-awgn}
\includegraphics{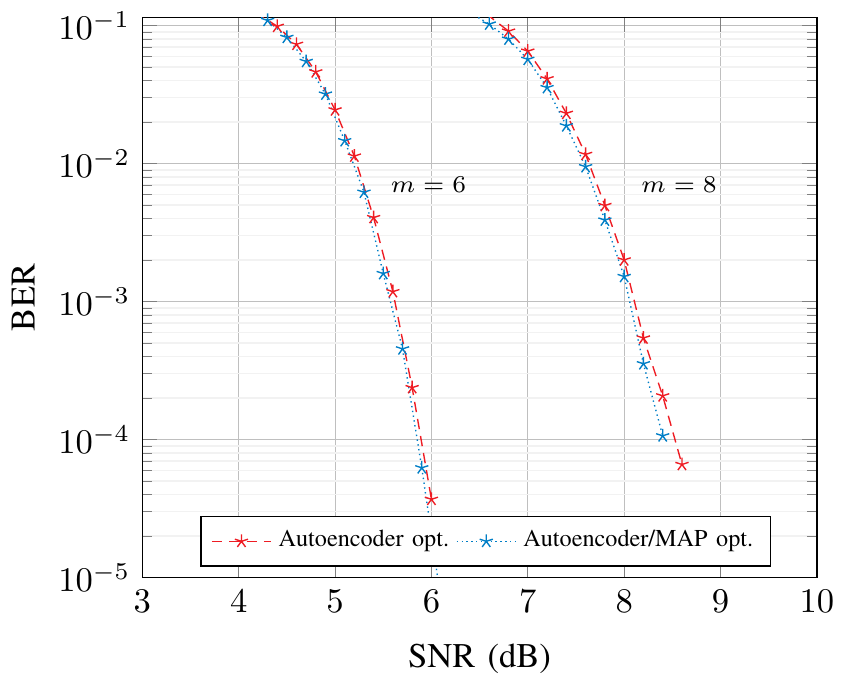}
	\subcaption{Simulated AWGN channel}\label{fig:ber-ota-map-awgn}
	\end{subfigure}
\begin{subfigure}[t]{\columnwidth}\centering
		\tikzsetnextfilename{ber-ota-map-ota}
\includegraphics{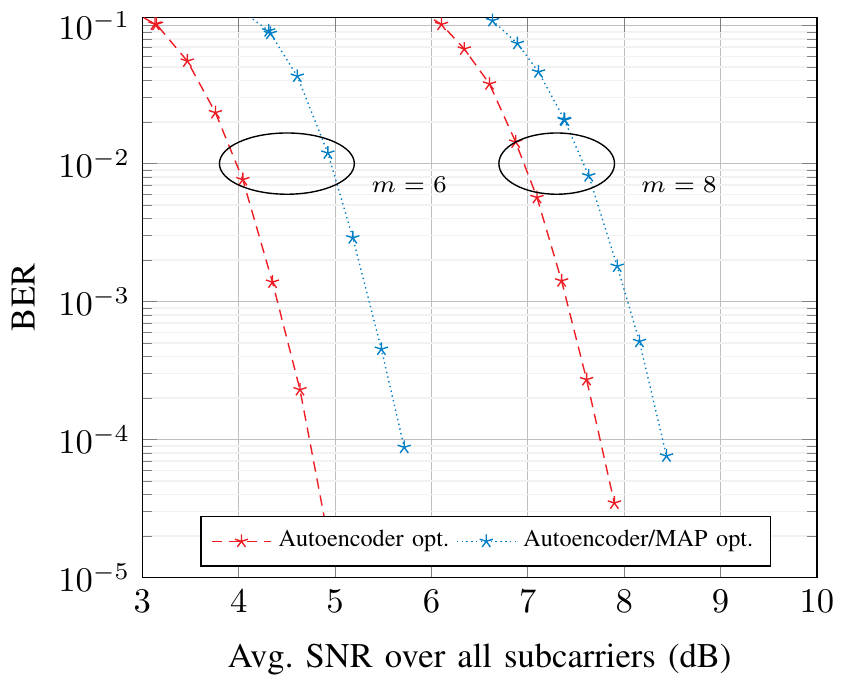}
				\subcaption{Actual over-the-air channel}\label{fig:ber-ota-map-ota}
	\end{subfigure}
	\caption{Comparison between autoencoder/MAP, using a \ac{MAP} demapper, and a pure autoencoder, using an NN-based (learned) demapper. The upper plot shows the \ac{BER} for the simulated \ac{AWGN} channel and the lower plot shows the \ac{OTA} \ac{BER} achieved  for $m=6$ and $m=8$. The SNR is averaged over all subcarriers of the OFDM system for the \ac{OTA} results. All schemes use IDD with 80 iterations.\label{fig:ber-ota-map}}
\end{figure}

%% file: tex/conclu.tex
We have described and experimentally validated training-based optimization of the physical layer of a point-to-point communication system, including an outer channel code. The key novelty is training on the bit-wise mutual information which enables seamless integration with  \ac{BMD} receivers, widely used in practice. The performance improvements over classic \ac{QAM} baselines result from geometric constellation shaping as well as learning of the optimal demapper. Further gains can be achieved through iterative demapping and decoding which has been elegantly integrated into the end-to-end learning procedure.
Once the system is trained, LDPC code optimization can lead to additional improvements.

Simulations as well as over-the-air experiments, in which the system was optimized from scratch over an actual wireless channel, have demonstrated the universality and practical benefits of the proposed method. The trained system outperforms a 256-\ac{QAM} baseline with an 802.11n \ac{LDPC} code in a realistic setup by approximately 1.3 dB while maintaining the same communication rate.

Looking into the future, we believe that learning-based optimization of the full physical layer for a point-to-point link
could be completely automated and may be one of the key ingredients of next generation communication systems. Due to the necessary overhead for training, such an approach seems particularly attractive for fixed wireless or optical channels. 
Interesting directions for further research comprise learning of probabilistic shaping \cite{GC2019} as well as schemes for multiuser communications, i.e., multiple access and broadcast channels \cite{stauffer2019deep}. Also, the idea of meta learning for fast online learning seems very promising \cite{park2019meta}.

%% file: tex/acknowledgment.tex
The authors would like to thank Laurent Schmalen for his help with the \ac{LDPC} code design algorithms and Marc Gauger for his help with the \ac{OTA} setup.